\documentclass[fleqn,usenatbib]{mnras}

\usepackage{newtxtext,newtxmath}

\usepackage[T1]{fontenc}

\DeclareRobustCommand{\VAN}[3]{#2}
\let\VANthebibliography\thebibliography
\def\thebibliography{\DeclareRobustCommand{\VAN}[3]{##3}\VANthebibliography}

\usepackage{graphicx}	
\usepackage{amsmath}	
\usepackage{amssymb}	
\usepackage{ae,aecompl}

\title[Thermal inertias of comet 67P/C--G]{Thermal inertias of pebble-pile comet 67P/Churyumov--Gerasimenko}

\author[Sota Arakawa and Kazumasa Ohno]{
Sota Arakawa$^{1,2}$\thanks{E-mail: sota.arakawa@nao.ac.jp}
and Kazumasa Ohno$^{2}$
\\
$^{1}$Division of Science, National Astronomical Observatory of Japan, Mitaka, Tokyo, 181-8588, Japan.\\
$^{2}$Department of Earth and Planetary Sciences, Tokyo Institute of Technology, Meguro, Tokyo, 152-8551, Japan.
}

\date{Accepted 2020 July 4. Received 2020 May 29; in original form 2020 May 29}

\pubyear{2020}

\begin{document}
\label{firstpage}
\pagerange{\pageref{firstpage}--\pageref{lastpage}}
\maketitle

\begin{abstract}
The {\it Rosetta} mission to comet 67P/Churyumov--Gerasimenko has provided new data to better understand what comets are made of.
The weak tensile strength of the cometary surface materials suggests that the comet is a hierarchical dust aggregate formed through gravitational collapse of a bound clump of small dust aggregates so-called ``pebbles'' in the gaseous solar nebula.
Since pebbles are the building blocks of comets, which are the survivors of planetesimals in the solar nebula, estimating the size of pebbles using a combination of thermal observations and numerical calculations is of great importance to understand the planet formation in the outer solar system.
In this study, we calculated the thermal inertias and thermal skin depths of the hierarchical aggregates of pebbles, for both diurnal and orbital variations of the temperature.
We found that the thermal inertias of the comet 67P/Churyumov--Gerasimenko are consistent with the hierarchical aggregate of cm- to dm-sized pebbles.
Our findings indicate that the icy planetesimals may have formed via accretion of cm- to dm-sized pebbles in the solar nebula.
\end{abstract}




\section{Introduction}

Comets are small and irregular-shaped objects composed of ice, organics, and refractory materials.
It is thought that they are formed in the outer region of the solar nebula, where the disk temperature is much lower than the sublimation temperature of ${\rm H}_{2}{\rm O}$ ice.
Given that comets spent a long time under cold conditions once they are formed, they are pristine objects and provide important clues about the environment of the early solar system.

The process by which micron-sized interstellar dust grains evolve into comets is still enigmatic.
In the context of planetesimal formation, the direct aggregation hypothesis was proposed to explain the origin of small icy bodies \citep[e.g.,][]{Okuzumi+2012}.
In this model, dust aggregates are transformed into km-sized comets via collisional growth and static compression \citep{Kataoka+2013b}, and the resulting comets are porous and homogeneous aggregates composed of $\mu$m-sized grains \citep[see also][]{Tsukamoto+2017,Homma+2018}.
In contrast, if comets are formed via gravitational collapse of a concentrated clump of mm- to dm-sized compressed dust aggregates called ``pebbles'' \citep[e.g.,][]{Johansen+2007b,Yang+2017}, then their internal structure would be described by ``hierarchical aggregates,'' i.e., loose agglomerates of pebbles \citep[e.g.,][]{Gundlach+2012,Skorov+2012,Blum+2017}.

The {\it Rosetta} mission to comet 67P/Churyumov--Gerasimenko (hereinafter referred to as comet 67P/C--G) has yielded a large amount of data for determining the internal structure of these objects.
Remarkably, the tensile strength of comet 67P/C--G was estimated from its surface topography, i.e., cliffs and overhangs \citep[e.g.,][]{Groussin+2015,Attree+2018}, and also based on crack propagation across the neck of the nucleus \citep{Hirabayashi+2016}.
The estimated tensile strength at the cometary surface is $ \lesssim 1\ {\rm Pa}$ \citep{Attree+2018}.
A low value of the tensile strength is also necessary to explain the dust activity of comets.
This is because typical gas pressures caused by the sublimation of ice (${\rm H}_{2}{\rm O}$, ${\rm C}{\rm O}_{2}$, and ${\rm C}{\rm O}$) beneath the covering dust layer may be on the order of $0.1$--$1\ {\rm Pa}$ \citep[e.g.,][]{Skorov+2012,Gundlach+2015}, and the sublimation gas pressure should exceed the tensile strength to drive dust activity.

The thermal and mechanical properties must be dependent on the structure of the dust aggregates, i.e., whether homogeneous or hierarchical \citep[see, e.g.,][]{Blum2018}.
\citet{Tatsuuma+2019} numerically investigated the tensile strength of homogeneous dust aggregates, $Y_{\rm hom}$, and they revealed that $Y_{\rm hom} > 10^{3}\ {\rm Pa}$ for homogeneous dust aggregates consisting of micron-sized monomer grains \citep[see also][]{Seizinger+2013b,Arakawa+2019e}.
Their numerical results are consistent with experimental results \citep[e.g.,][]{Blum+2006}; however, the obtained tensile strength substantially exceeds the maximum sublimation pressure of ice at the cometary surface.
In contrast, \citet{Blum+2014} experimentally measured the tensile strength of hierarchical aggregates of millimetre-sized pebbles, and they found that the tensile strength of the uncompressed hierarchical aggregates, $Y_{\rm hie, 0}$, is on the order of $0.1$--$1\ {\rm Pa}$.
The dust activity of comets can then be driven by the sublimation of ice if comets are hierarchical aggregates of pebbles.
The tensile strength of compressed hierarchical aggregates, $Y_{\rm hie}$, is given by $Y_{\rm hie} \sim 0.03 p + Y_{\rm hie, 0}$, where $p$ is the compression pressure before breaking up \citep[see][]{Blum+2014}.
The volume-averaged pressure of the cometary interior is $35\ {\rm Pa}$ for the larger lobe of comet 67P/C--G \citep{Blum+2017}, and the obtained tensile strength, $Y_{\rm hie} \sim 1\ {\rm Pa}$, is also consistent with the estimates from the {\it Rosetta} mission \citep[e.g.,][]{Groussin+2015,Attree+2018}.

As previously indicated, the tensile strength of comet 67P/C--G is consistent with the hierarchical aggregate model proposed by \citet{Skorov+2012}.
The compressive strength of the surface material of the comet can also be reproduced by the hierarchical aggregate model \citep[see][]{Heinisch+2019}.
Therefore, the gravitational collapse of a concentrated clump of pebbles in the solar nebula is the best model that explains the formation process of comets.
However, the size of the pebbles was poorly constrained in previous studies.

Heat transport via the surface of the nucleus is a fundamental process of comets.
It is mainly driven by solar illumination, and the diurnal and orbital variations of the energy flux cause temperature variations of the surface layer.
Thermal inertia and thermal skin depth are the key parameters involved in the propagation of energy into the cometary interior \citep[although surface roughness also plays an important role in the heat transfer process; e.g.,][]{Marshall+2018}.
Since thermal inertia reflects the size and porosity of regolith and boulders on the surface of small bodies \citep[e.g.,][]{Okada+2017,Okada+2020}, we could apply constraints on the size of the pebbles from the thermal inertia of comet 67P/C--G.

In this study, we calculate the thermal inertia of comet 67P/C--G for both diurnal and orbital temperature variations and discuss the dependence of the thermal inertia on the pebble radius.
In Section \ref{sec.2}, we describe the models of dust aggregates used in this study.                                                                                                                                                                                                                                                                                                                                                                           
In Section \ref{sec.3}, we present numerical results for the diurnal and orbital thermal inertias of comet 67P/C--G and compare our calculations with observational results.
We found that the observed thermal inertias are consistent with the hierarchical aggregate model when the pebbles are cm-sized or larger aggregates.
In contrast, the thermal inertias of hierarchical aggregates composed of mm-sized or smaller pebbles are too low to explain the observed thermal inertias.
We briefly highlight the constraint on the size of the pebbles in the literature in Section \ref{sec.5}, and a summary is presented in Section \ref{sec.6}.

\section{Modeling of Dust Aggregates}
\label{sec.2}

In Section \ref{sec.2}, we describe the model of hierarchical aggregates used in this study.
We introduce a core--mantle monomer grain model in Section \ref{sec.monomer}. 
In Section \ref{sec.hie}, we briefly review the idea of hierarchical aggregates proposed by \citet{Skorov+2012}.
In Section \ref{sec.density}, we discuss the material composition of comet 67P/C--G. 
Finally, we explain the thermal properties of dust aggregates in Section \ref{sec.thermal} \citep[see also][]{Arakawa+2019a}.

\subsection{Monomer grains}
\label{sec.monomer}

\begin{figure}
\centering
\includegraphics[width = 0.8\columnwidth]{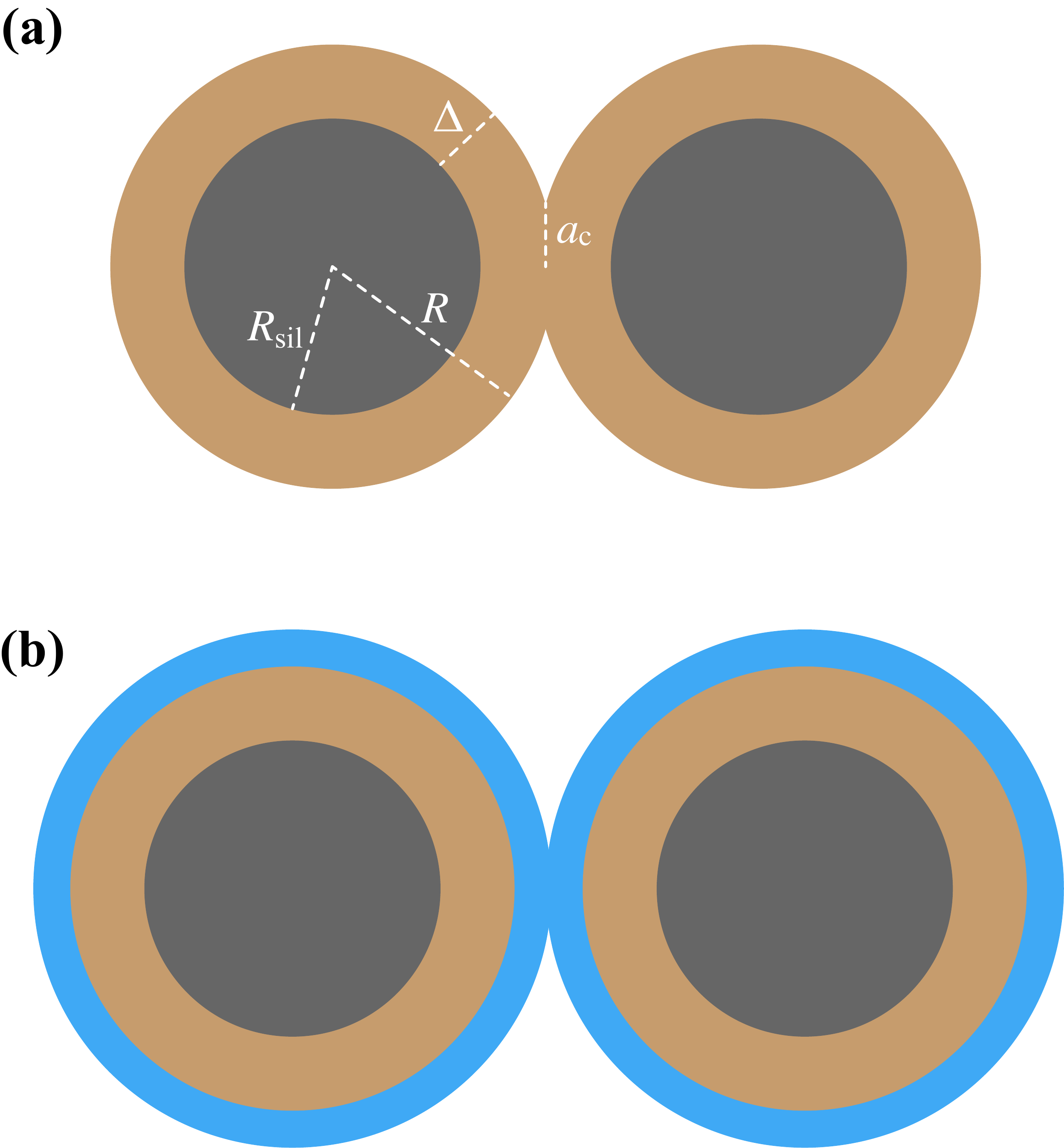}
\caption{
Schematic illustration of two contacting monomer grains with a radius of $R$.
(a) The case of organic--silicate grains.
(b) The case of ice--organic--silicate grains.
The radius of the silicate core is $R_{\rm sil} = 0.5\ {\mu}{\rm m}$ in both cases (a) and (b).
When the contact radius, $a_{\rm c}$, is smaller than the thickness of the outermost layer, $\Delta$, the inner layers have no effect on the adhesion of the monomer grains.
}
\label{fig1}
\end{figure}

In this study, we assume that monomer grains have a core--mantle structure \citep[e.g.,][]{Homma+2019}.
The {\it Rosetta} mission revealed that cometary dust particles ejected from the surface of comet 67P/C--G are a mixture of anhydrous silicates and organics \citep[e.g.,][]{Bardyn+2017}.
Organic materials also exist in chondritic porous interplanetary dust particles \citep[IDPs; e.g.,][]{Flynn+2013}.
These chondritic porous IDPs have a cometary origin.
They represent the pristine materials in the solar nebula \cite[e.g.,][]{Ishii+2008}, and individual $\mu$m-sized grains are mantled by organics \citep[e.g.,][]{Flynn+2013}.
Based on these facts, we consider silicate grains coated by organic mantles (organic--silicate grains, see case (a) of Figure \ref{fig1}). 
In addition, we also consider the ice--organic--silicate grains (see case (b) of Figure \ref{fig1}) because comets retain ice in their subsurface region.

A model for two adhered homogeneous and spherical grains was proposed by \citet*{Johnson+1971}, called JKR contact theory \citep[see also][]{Johnson1987,Dominik+1997,Wada+2007}.
In JKR theory, the contact radius of two adhered spherical monomers, $a_{\rm c}$, is given by
\begin{equation}
a_{\rm c} = {\left[ \frac{9 \pi \gamma {\left( 1 - \nu^{2} \right)}}{2 E R} \right]}^{1/3} R,
\label{eqac}
\end{equation}
where $\gamma$ is the surface energy, $E$ is Young's modulus, $\nu$ is the Poisson's ratio, and $R$ is the monomer radius.
We summarize the material properties adopted in this study in Appendix \ref{appendix:A}.

The stress distribution in contacting monomers around the contact area is given in \citet{Johnson1987}, and the spatial scale of the stress distribution is $a_{\rm c}$.
Therefore, for the case of two contacting core-mantle grains, the contact radius is determined by the material properties of the outermost layer when its thickness, $\Delta$, is larger than the contact radius, $a_{\rm c}$. 
Schematic figures of two core--mantle grains in contact are shown in Figure \ref{fig1}.

\subsubsection{Organic--silicate grains}

In this study, we set the radius of the silicate core as $R_{\rm sil} = 0.5\ {\mu}{\rm m}$, which is consistent with the size of monomer particles reported by the {\it Rosetta} mission \citep{Bentley+2016,Mannel+2016}.
The mass fractions of the organic mantle and silicate core are $f_{\rm org}$ and $f_{\rm sil}$, and the volume fractions of the organic mantle and silicate core are given by
\begin{equation}
\chi_{\rm org} = \frac{{f_{\rm org} / \rho_{\rm org}}}{{f_{\rm org} / \rho_{\rm org}} + {f_{\rm sil} / \rho_{\rm sil}}},
\end{equation}
and 
\begin{equation}
\chi_{\rm sil} = \frac{{f_{\rm sil} / \rho_{\rm sil}}}{{f_{\rm org} / \rho_{\rm org}} + {f_{\rm sil} / \rho_{\rm sil}}},
\end{equation}
respectively, where $\rho_{\rm org}$ and $\rho_{\rm sil}$ are the material density of organic and silicate.
The monomer radius of the organic--silicate grains is given by
\begin{equation}
R = {\chi_{\rm sil}}^{- 1/3} R_{\rm sil},
\end{equation}
and the thickness of the organic mantle is therefore given by
\begin{equation}
\Delta = {\left( 1 - {\chi_{\rm sil}}^{1/3} \right)} R.
\end{equation}
The grain density of organic--silicate grains is given by
\begin{equation}
\rho_{\rm m} = \chi_{\rm org} \rho_{\rm org} + \chi_{\rm sil} \rho_{\rm sil}.
\end{equation}
The material properties used in this study are listed in Table \ref{table1}.

\subsubsection{Ice-organic--silicate grains}

We can obtain the monomer radius and the thickness of the ice mantle of ice--organic--silicate grains in a similar manner to the organic--silicate grains.
The mass fractions of the ice shell, organic mantle, and silicate core are $f_{\rm ice}$, $f_{\rm org}$, and $f_{\rm sil}$, respectively.
The volume fractions of the ice shell, organic mantle, and silicate core are given by
\begin{equation}
\chi_{\rm ice} = \frac{{f_{\rm ice} / \rho_{\rm ice}}}{{f_{\rm ice} / \rho_{\rm ice}} + {f_{\rm org} / \rho_{\rm org}} + {f_{\rm sil} / \rho_{\rm sil}}},
\end{equation}
\begin{equation}
\chi_{\rm org} = \frac{{f_{\rm org} / \rho_{\rm org}}}{{f_{\rm ice} / \rho_{\rm ice}} + {f_{\rm org} / \rho_{\rm org}} + {f_{\rm sil} / \rho_{\rm sil}}},
\end{equation}
and 
\begin{equation}
\chi_{\rm sil} = \frac{{f_{\rm sil} / \rho_{\rm sil}}}{{f_{\rm ice} / \rho_{\rm ice}} + {f_{\rm org} / \rho_{\rm org}} + {f_{\rm sil} / \rho_{\rm sil}}},
\end{equation}
respectively.
Then, the monomer radius of ice--organic--silicate grains is given by $R = {\chi_{\rm sil}}^{- 1/3} R_{\rm sil}$, and the thickness of the ice shell is therefore given by
\begin{equation}
\Delta = {\left[ 1 - {\left( \chi_{\rm org} + \chi_{\rm sil} \right)}^{1/3} \right]} R.
\end{equation}
The grain density of the ice--organic--silicate grains is
\begin{equation}
\rho_{\rm m} = \chi_{\rm ice} \rho_{\rm ice} + \chi_{\rm org} \rho_{\rm org} + \chi_{\rm sil} \rho_{\rm sil}.
\end{equation}

\subsection{Hierarchical aggregate}
\label{sec.hie}

\begin{figure}
\centering
\includegraphics[width = \columnwidth]{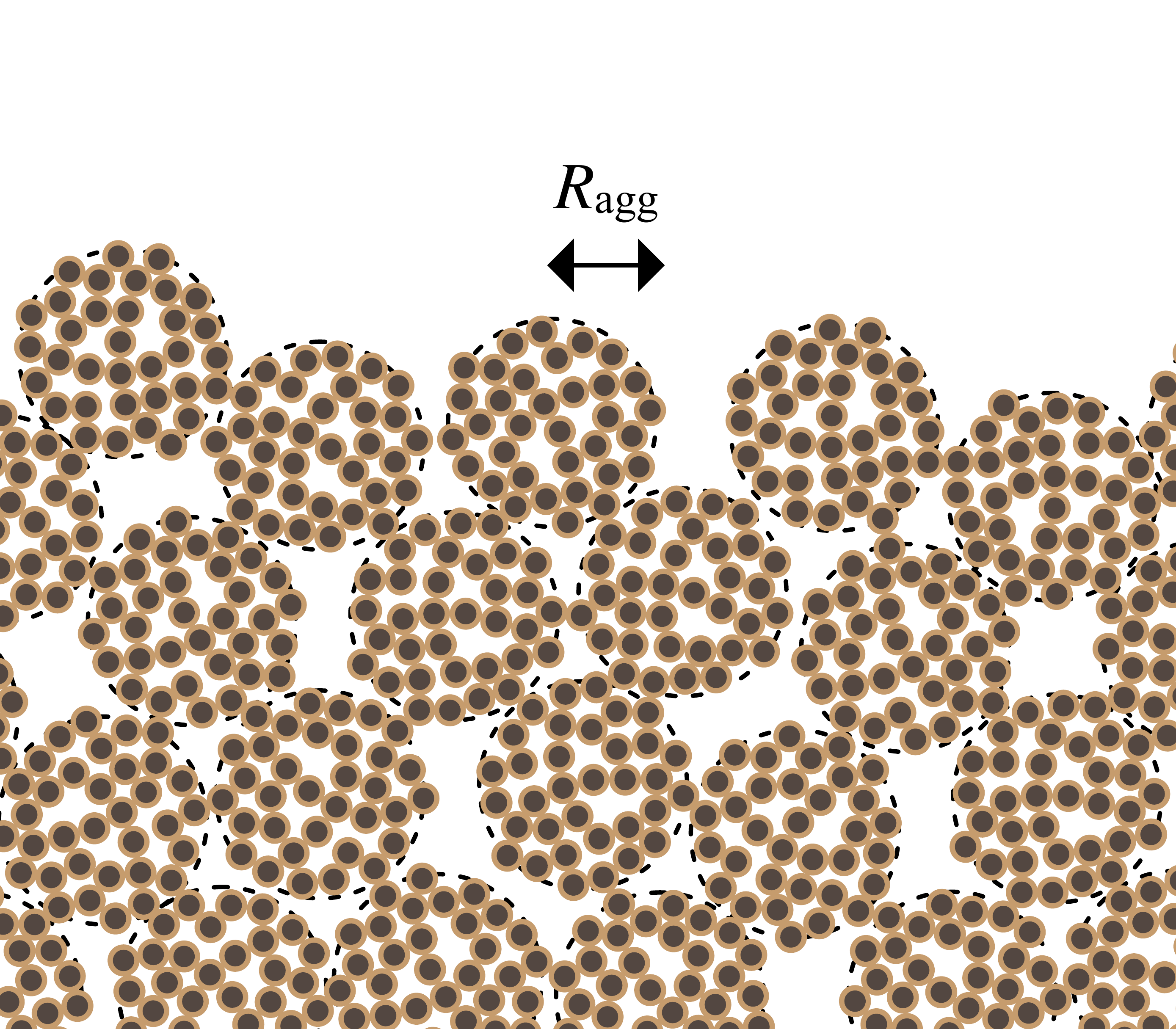}
\caption{
Schematic illustration of a hierarchical aggregate.
A hierarchical aggregate formed via accretion of the constituent aggregates (surrounded by dashed circles).
The aggregate radius of the constituent aggregates (i.e., ``pebbles'') is $R_{\rm agg}$ and the monomer radius is $R$.
The filling factor of the constituent aggregates is $\phi_{\rm agg}$, and the filling factor of the aggregate packing structure is $\phi_{\rm p}$.
Thus, the total filling factor of the hierarchical aggregate is given by $\phi_{\rm total} = \phi_{\rm agg} \phi_{\rm p}$.
}
\label{fig2}
\end{figure}

Figure \ref{fig2} shows schematic illustrations of a hierarchical aggregate \citep[see also][]{Gundlach+2012,Skorov+2012}.
If comet nuclei are formed via gravitational collapse of a bound clump of pebbles, their packing morphology would as shown in Figure \ref{fig2}.
The concept of the hierarchical aggregate model is described in Section 2 of \citet{Skorov+2012} and also in Section 2 of \citet{Blum+2017}.
We briefly summarize the scenario for comet formation in the following sections.

\subsubsection{Formation of pebbles via collisions}

The first step of planetesimal formation is the collisional growth of dust particles in the gaseous solar nebula \citep[][and references therein]{Blum+2008}.
Aggregates initially collide at very low speeds, which results in the growth of the aggregates until their size reaches the bouncing and/or fragmentation barriers \citep{Brauer+2008,Zsom+2010}.
Depending on the solar nebula model, the threshold size of the barriers is in the range of (sub)millimetres to metres \citep{Skorov+2012}.

Continued non-sticking collisions lead to rounding and compaction of the aggregates \citep{Weidling+2009,Weidling+2012}.
Whether adhesion collisions occur or not mainly depends on the filling factor of the aggregates, $\phi_{\rm agg}$.
When the filling factor is higher than $0.1$, sticking collisions are infrequently observed in laboratory experiments \citep[e.g.,][]{Langkowski+2008}.
The filling factor of the aggregates then approaches $0.35$--$0.4$ as a consequence of mutual collisions \citep[e.g.,][]{Weidling+2009,Guettler+2010}.

\subsubsection{Formation of planetesimals via gravitational instability}

\citet{Johansen+2007b} proposed that planetesimal formation occurred via spatial concentration of pebbles due to streaming instability \citep[e.g.,][]{Youdin+2005,Johansen+2007a}.
Streaming instability leads to the formation of a gravitationally bound cloud of pebbles in the solar nebula, which gently collapses to form planetesimals.

Spontaneous concentration of pebbles due to streaming instability can occur when the Stokes number of the pebbles, ${\rm St}$, is in the range $10^{-3} \lesssim {\rm St} \lesssim 1$ \citep{Carrera+2015,Yang+2017}.
The Stokes number is defined as ${\rm St} = t_{\rm s} \Omega_{\rm K}$, where $t_{\rm s}$ is the stopping time of the pebbles and $\Omega_{\rm K}$ is the Kepler frequency.
Assuming a minimum mass solar nebula model \citep{Weidenschilling1977b,Hayashi1981}, compressed pebbles with radii in the range of $0.1\ {\rm mm}$ and $10\ {\rm cm}$ can be concentrated due to streaming instability if comets form at $\sim 10$--$30\ {\rm au}$ from the Sun \citep[see][]{Blum+2017}.

We note that other mechanisms can account for the spatial concentration of dust aggregates in the gaseous solar nebula, e.g., dust trapping at the local pressure maxima \citep[e.g.,][]{Haghighipour+2003} and the vortices generated by hydrodynamical instabilities \citep[e.g.,][]{Meheut+2012}.
A wide range of aggregate sizes, from (sub)mm- to metre-sized, could concentrate in the turbulent solar nebula \citep[][and references therein]{Johansen+2014}.
Therefore, the pebbles, which are the building blocks of comets, would also be (sub)mm- to metre-sized dust aggregates formed in the solar nebula.

The concentration of pebbles using a gentle gravitational collapse results in the formation of comets with a filling factor of the aggregate packing structure of $\phi_{\rm p} \sim \phi_{\rm RCP}$, where $\phi_{\rm RCP} = 0.64$ is the filling factor for random close packing \citep[e.g.,][]{Berryman1983}.
The total filling factor of the hierarchical aggregate, $\phi_{\rm total} = \phi_{\rm agg} \phi_{\rm p}$, is approximately $0.22$--$0.26$, which is compatible with the estimates obtained from the {\it Rosetta} mission \citep[e.g.,][]{Kofman+2015,Paetzold+2016}.

\subsection{Material density and mass fraction}
\label{sec.density}

In this section, we discuss the material density of cometary organics and silicates, and we also evaluate the mass fraction of ice, organics, and silicates.

\subsubsection{Material density of ${\rm H}_{2}{\rm O}$ ice}

The material density of crystalline ${\rm H}_{2}{\rm O}$ ice is $\rho_{\rm ice} = 920\ {\rm kg}\ {\rm m}^{-3}$.
We note that the material density of amorphous ${\rm H}_{2}{\rm O}$ ice is $940\ {\rm kg}\ {\rm m}^{-3}$ \citep{Mishima+1985} and the difference between crystalline and amorphous ice is small.

\subsubsection{Material density of cometary organics}

There are some analogues for cometary organics, e.g., macromolecular insoluble organic matter (IOM), HCN heteropolymers, and bitumen.
We estimated the material density of the organics, $\rho_{\rm org}$, based on these analogues.

The elemental composition of organic matter in cometary dust is essentially chondritic and shares similarities with macromolecular IOM in carbonaceous chondrites \citep{Fray+2016}.
The aliphatic signatures in the infrared spectrum of comet 67P/C--G are also compatible with those of carbonaceous chondrites \citep{Raponi+2020}.
The material density of IOM is in the range of $1200$--$1400\ {\rm kg}\ {\rm m}^{-3}$ \citep{Zolotov2020}.
The material density of an HCN heteropolymer, a reasonable candidate for the dark lag deposit of cometary nuclei, is $1620\ {\rm kg}\ {\rm m}^{-3}$ \citep{Khare+1994}.
Natural solid oil bitumen is also a useful spectral analogue for cometary refractory organics, and its material density is in the range of $1050$--$2000\ {\rm kg}\ {\rm m}^{-3}$ \citep[see][and references therein]{Moroz+1998}.
Therefore, a reasonable range for the material density of cometary organics is
\begin{equation}
1000\ {\rm kg}\ {\rm m}^{-3} \leq \rho_{\rm org} \leq 2000\ {\rm kg}\ {\rm m}^{-3}.
\end{equation}

\subsubsection{Material density of silicate}

The grain density of carbonaceous chondrites was reported by \citet{Consolmagno+2008}: $2460\ {\rm kg}\ {\rm m}^{-3}$ for CI chondrites, $2900\ {\rm kg}\ {\rm m}^{-3}$ for CM chondrites, and $3580\ {\rm kg}\ {\rm m}^{-3}$ for CK chondrites.
\footnote{
The carbon content of carbonaceous chondrites is $\sim 1 \%$ \citep[e.g.,][]{Gail+2017}, and the presence of organics hardly modifies the grain density of carbonaceous chondrites.
}
The higher density carbonaceous chondrites (e.g., CK) are anhydrous whereas the lower density carbonaceous chondrites (CI and CM) are hydrated.
We assume that the material density of a silicate is,
\begin{equation}
\rho_{\rm sil} = 2500\ {\rm kg}\ {\rm m}^{-3}\ {\rm or}\ 3500\ {\rm kg}\ {\rm m}^{-3}.
\end{equation}

\subsubsection{Refractory-to-ice mass ratio}

The refractory-to-ice mass ratio in the nucleus,
\begin{align}
\delta_{\rm RI} &\equiv \frac{f_{\rm sil} + f_{\rm org}}{f_{\rm ice}}, \nonumber \\
                &= \frac{1}{f_{\rm ice}} - 1,
\end{align}
has been estimated based on several studies \citep[e.g.,][]{Fulle+2017,Fulle+2019a,Paetzold+2019}.
Considering the average dust bulk density of the dust particles ejected from the nucleus that were collected during the entire mission, \citet{Fulle+2017} obtained the refractory-to-ice mass ratio $\delta_{\rm RI} = 7.5$ inside the nucleus.
\citet{Fulle+2019a} also estimated $\delta_{\rm RI}$ inside the nucleus from the mass balance considering dust loss, water loss (both from the nucleus and distributed sources) and dust fallout.
Assuming that the dust-to-gas mass ratio in the lost material is in the range of $0.7$ and $2.0$, the refractory-to-ice mass ratio is in the range of $4.3 < \delta_{\rm RI} < 55$.
\citet{Paetzold+2019} discussed the range of $\delta_{\rm RI}$ inside the nucleus that is compatible with the bulk density, and the suggested range is $3 < \delta_{\rm RI} < 7$.

Therefore, we conclude that the possible range of the refractory-to-ice mass ratio inside the nucleus is $3 \lesssim \delta_{\rm RI} \lesssim 10$, and in this section, we assume that the mass fraction of ice is,
\begin{equation}
f_{\rm ice} = 0.1\ {\rm or}\ 0.2,
\end{equation} 
which corresponds to $\delta_{\rm RI} = 9$ and $4$, respectively.
We note that \citet{Lorek+2016} also suggested that the refractory-to-ice mass ratio should be $3 \lesssim \delta_{\rm RI} \lesssim 9$ based on the results of Monte Carlo simulations of collisional evolution of pebbles.
Our evaluation of $f_{\rm ice}$ is consistent with the results of \citet{Lorek+2016}.

\subsubsection{The mass fraction of organics in refractory dust grains}

The mass fraction of organics in refractory dust grains, $\delta_{\rm OR} \equiv f_{\rm org} / {( f_{\rm sil} + f_{\rm org} )}$, has also been estimated in several studies \citep[e.g.,][]{Bardyn+2017,Fulle+2018}.
\citet{Fulle+2018} concluded that the Grain Impact Analyser and Dust Accumulator \citep[GIADA;][]{Colangeli+2007} observed average organic mass fractions of $\delta_{\rm OR} = 38 \pm 8 \%$.
The mass fraction of organics was also measured using the Cometary Secondary Ion Mass Analyser \citep[COSIMA;][]{Kissel+2007}, and \citet{Bardyn+2017} found that the mass fraction of organics is $\delta_{\rm OR} \sim 45 \pm 15 \%$, which is consistent with the result of \citet{Fulle+2018}.

We assume $f_{\rm org} = 0.3$ in Section \ref{sec.bulk}.
The resulting organic mass fraction is $\delta_{\rm OR} = 33.33... \%$ for the case of $f_{\rm ice} = 0.1$ and $\delta_{\rm OR} = 37.5 \%$ for the case of $f_{\rm ice} = 0.2$, respectively.

\subsubsection{Constraint on material density and mass fraction based on the bulk density of comet 67P/C--G}
\label{sec.bulk}

The bulk density of comet 67P/C--G, $\rho_{\rm bulk} = 533\ {\rm kg}\ {\rm m}^{-3}$ \citep{Paetzold+2016}, is given by
\begin{align}
\rho_{\rm bulk} &= \phi_{\rm agg} \phi_{\rm p} {\left( \chi_{\rm sil} \rho_{\rm sil} + \chi_{\rm org} \rho_{\rm org} + \chi_{\rm ice} \rho_{\rm ice} \right)}, \nonumber \\
                &= \frac{\phi_{\rm agg} \phi_{\rm p}}{{f_{\rm ice} / \rho_{\rm ice}} + {f_{\rm org} / \rho_{\rm org}} + {f_{\rm sil} / \rho_{\rm sil}}}.
\label{eq.bulk}
\end{align}
We can then obtain the parameter range of the material density values, $\rho_{\rm org}$ and $\rho_{\rm sil}$, and the mass fraction of ice, $f_{\rm ice}$, by solving Equation (\ref{eq.bulk}).

\begin{figure*}
\centering
\includegraphics[width = \columnwidth]{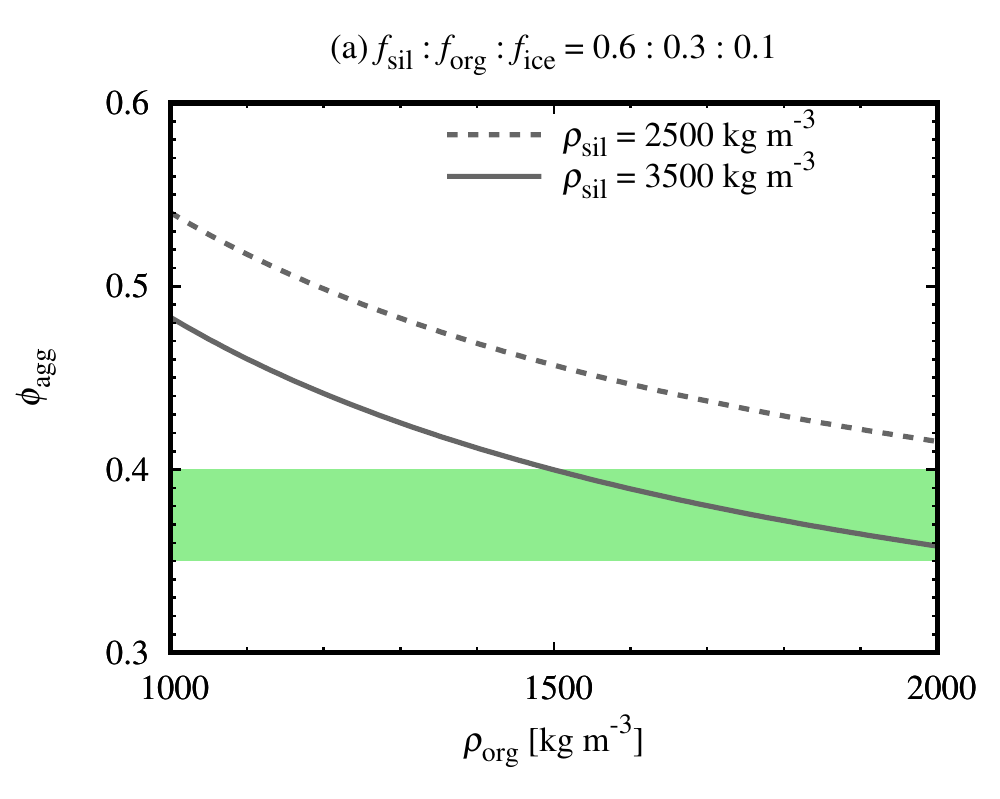}
\includegraphics[width = \columnwidth]{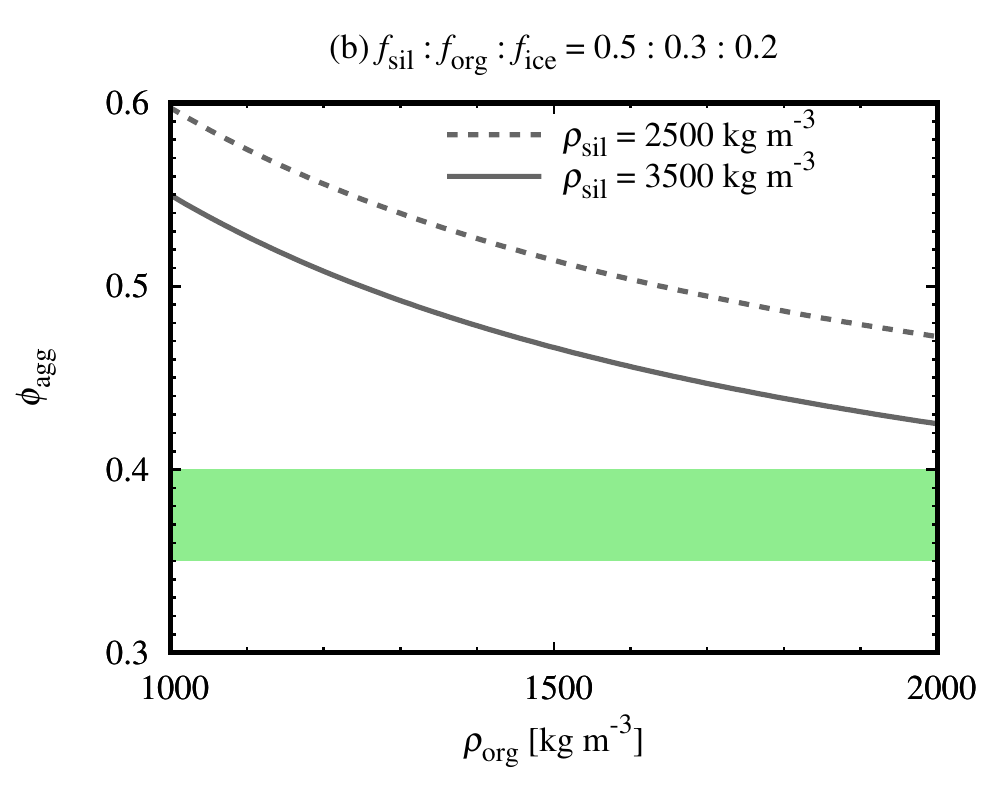}
\caption{
Filling factor of constituent aggregates, $\phi_{\rm agg}$, and the material density of the organics, $\rho_{\rm org}$, are calculated from the bulk density of comet 67P/C--G using Equation (\ref{eq.bulk}).
(a) For the case of $f_{\rm sil} : f_{\rm org} : f_{\rm ice} = 0.6 : 0.3 : 0.1$.
(b) For the case o $f_{\rm sil} : f_{\rm org} : f_{\rm ice} = 0.5 : 0.3 : 0.2$.
The dashed lines represent the case of $\rho_{\rm sil} = 2500\ {\rm kg}\ {\rm m}^{-3}$ (hydrated carbonaceous chondrites) whereas the solid lines represent the case of $\rho_{\rm sil} = 3500\ {\rm kg}\ {\rm m}^{-3}$ (anhydrous carbonaceous chondrites).
The green shaded regions represent the range of $\phi_{\rm agg}$ when pebbles are formed via multiple non-sticking collisions \citep{Weidling+2009,Guettler+2010}.
}
\label{fig3}
\end{figure*}

Figure \ref{fig3} shows the filling factor of the constituent aggregates, $\phi_{\rm agg}$, as a function of the material density of the organics, $\rho_{\rm org}$.
In the framework of the hierarchical aggregate model, the filling factor of the aggregate packing structure is $\phi_{\rm p} = 0.64$ and the filling factor of the constituent aggregates is $\phi_{\rm agg} \simeq 0.35$--$0.4$ (green shaded regions).
As shown in Figure \ref{fig3}(a), we can reproduce the bulk density of comet 67P/C--G when $f_{\rm ice} = 0.1$, $\rho_{\rm sil} = 3500\ {\rm kg}\ {\rm m}^{-3}$ and $\rho_{\rm org} \gtrsim 1500\ {\rm kg}\ {\rm m}^{-3}$.
The value of $\rho_{\rm sil} = 3500\ {\rm kg}\ {\rm m}^{-3}$ is consistent with the fact that the mineral phase in dust grains measured using COSIMA is predominantly composed of anhydrous silicates \citep{Bardyn+2017}.

In contrast, when we assume $f_{\rm ice} = 0.2$, we cannot reproduce the bulk density of comet 67P/C--G, even if the material density of the organics is $\rho_{\rm org} = 2000\ {\rm kg}\ {\rm m}^{-3}$, as shown in Figure \ref{fig3}(b).
We conclude that $f_{\rm ice} < 0.2$ (i.e., $\delta_{\rm RI} > 4$) is suitable for the hierarchical aggregate model from the perspective of the bulk density constraint.

In the rest of this paper, we set $\phi_{\rm agg} = 0.4$, $\rho_{\rm sil} = 3500\ {\rm kg}\ {\rm m}^{-3}$, and $\rho_{\rm org} = 1500\ {\rm kg}\ {\rm m}^{-3}$.
We also set the mass fractions of the organic--silicate grains and ice--organic--silicate grains as $f_{\rm org} : f_{\rm sil} = 1/3 : 2/3$ and $f_{\rm ice} : f_{\rm org} : f_{\rm sil} = 0.1 : 0.3 : 0.6$, respectively (see Tables \ref{table1} and \ref{table2}).
Assuming these parameters, the condition for using JKR theory, $a_{\rm c} < \Delta$, is satisfied for both organic--silicate grains and ice--organic--silicate grains.

\subsection{Thermal conductivity and specific heat capacity}
\label{sec.thermal}

\subsubsection{Thermal conductivity of constituent aggregates}
\label{sec.k.agg}

The thermal conductivity of tbe constituent aggregates, $k_{\rm agg}$, is dominated by the thermal conductivity through the solid network, $k_{\rm sol}$ (see Appendix \ref{app.rad}).
\citet{Arakawa+2017} obtained that $k_{\rm sol}$ is given by
\begin{equation}
k_{\rm sol} = 2 k_{\rm mat} \frac{a_{\rm c}}{R} f {\left( \phi_{\rm agg} \right)},
\end{equation}
where $f$ is the dimensionless (normalized) thermal conductivity and $k_{\rm mat}$ is the material thermal conductivity.
The dimensionless function $f$ depends on $\phi_{\rm agg}$ and the average coordination number $Z$; and $Z$ also depend on $\phi_{\rm agg}$.
Numerical simulations performed by \citet{Arakawa+2019a} revealed that $f$ and $Z$ are given by
\begin{equation}
f {\left( \phi_{\rm agg} \right)} = 0.784 {\phi_{\rm agg}}^{1.99} {\left( \frac{Z}{2} \right)}^{0.556},
\end{equation}
and
\begin{equation}
Z = 2 + 9.38 {\phi_{\rm agg}}^{1.62}.
\end{equation}
The physical backgrounds of these equations are described in \citet{Arakawa+2019e}.
We set $k_{\rm agg} = k_{\rm sol}$ in this study.

Heat flows through the monomer--monomer contacts, and the heat conductance at the contact determines the heat flow within two monomers.
A contact between two monomers disturbs the temperature profiles inside the grains only for the spatial scale of $a_{\rm c}$, as in the case of the stress distribution described in Section \ref{sec.monomer} \citep[see also][]{Gusarov+2003}.
For the case of core--mantle monomers, the material thermal conductivity of the outermost layer determines the thermal conductivity through the solid network when $a_{\rm c} < \Delta$ is satisfied.
We summarize the material thermal conductivities in Appendix \ref{appendix:B} (see Figure \ref{fig.k_T}).

\subsubsection{Thermal conductivity of a hierarchical aggregate}
\label{sec.k.hie}

In contrast, the thermal conductivity of hierarchical aggregates is dominated by radiative transfer within inter-aggregate voids \citep[e.g.,][]{Gundlach+2012}.

The thermal conductivity of hierarchical aggregates, $k_{\rm hie}$, is given by
\begin{equation}
k_{\rm hie} = \frac{16}{3} \sigma_{\rm SB} T^{3} l_{\rm mfp, hie},
\end{equation}
where $\sigma_{\rm SB}$ is the Stefan--Boltzmann constant, $T$ is the temperature, and $l_{\rm mfp, hie}$ is the mean free path of photons within the inter-aggregate structure of the hierarchical aggregates.
\footnote{
We note that the thermal conductivity of pebbles, $k_{\rm agg}$, may have an important effect on $k_{\rm hie}$ when $k_{\rm agg} < k_{\rm hie}$, due to the non-isothermality in each pebble \citep[see][]{Ryan+2020}.
}

In the same way as \citet{Gundlach+2012}, we can also evaluate the thermal conductivity through the solid network of hierarchical aggregates, $k_{\rm hie, sol}$, as follows:
\begin{equation}
k_{\rm hie, sol} = 2 k_{\rm agg} \frac{a_{\rm c, agg}}{R_{\rm agg}} f {\left( \phi_{\rm p} \right)},
\end{equation}
where $a_{\rm c, agg}$ is the contact radius of two adhered pebbles.
We note, however, that \citet{Gundlach+2012} revealed that $k_{\rm hie, sol}$ is negligibly small compared to $k_{\rm hie}$ when the size of the pebbles is larger than $0.1\ {\rm mm}$ \citep[see Figure 15 of][]{Gundlach+2012}.
This is because $a_{\rm c, agg} / R_{\rm agg}$ of the pebbles is much smaller than unity and heat transfer through the solid network is limited by the contact area between two adhered pebbles.
Therefore, we assume that the thermal conductivity of hierarchical aggregates is given by radiative transfer within inter-aggregate voids in this study.

We confirmed that the effective absorption cross-section of the constituent aggregates, $\sigma_{\rm eff, agg}$, is approximately equal to the geometric cross section, $\sigma_{\rm agg} = \pi {R_{\rm agg}}^{2}$, when $R_{\rm agg} \geq 0.1\ {\rm mm}$ (see Appendix \ref{app.sigma}).
The mean free path $l_{\rm mfp, hie}$ is then given by the following geometric optical approximation:
\begin{align}
l_{\rm mfp, hie} &= {\left( 1 - \phi_{\rm p} \right)} \frac{n_{\rm agg}}{\sigma_{\rm agg}}, \nonumber \\
                 &= \frac{4}{3} \frac{1 - \phi_{\rm p}}{\phi_{\rm p}} R_{\rm agg},
\label{eq.mfp}
\end{align}
where 
\begin{equation}
n_{\rm agg} = \frac{\phi_{\rm p}}{{\left( 4 \pi / 3 \right)} {R_{\rm agg}}^{3}},
\end{equation}
is the number density of the constituent aggregates.
\footnote{
In this case we assume that radiative heat transfer only occurs in the inter-aggregate voids and neglect the radiative heat transport inside the constituent aggregates \citep[see][]{Gundlach+2012}. 
}
Equation (\ref{eq.mfp}) exhibits excellent agreement with the empirical formula reported by \citet{Gundlach+2012}: $l_{\rm mfp, hie} \simeq 1.34 {\left[ {\left( 1 - \phi_{\rm p} \right)} / \phi_{\rm p} \right]} R_{\rm agg}$.
The typical distance among the constituent aggregates, $l_{\rm agg}$, is also given by
\begin{equation}
l_{\rm agg} = \frac{4 R_{\rm agg}}{3 \phi_{\rm p}}.
\end{equation}

\subsubsection{Specific heat capacity}

For the case of organic--silicate grains, the specific heat capacity of a monomer, $c_{\rm m}$, is given by
\begin{equation}
c_{\rm m} = f_{\rm org} c_{\rm org} + f_{\rm sil} c_{\rm sil},
\end{equation}
and for the case of ice--organic--silicate grains,
\begin{equation}
c_{\rm m} = f_{\rm ice} c_{\rm ice} + f_{\rm org} c_{\rm org} + f_{\rm sil} c_{\rm sil},
\end{equation}
where $c_{\rm ice}$, $c_{\rm org}$, and $c_{\rm sil}$ are the material specific heat capacities.
The specific heat capacities used in this study are summarized in Appendix \ref{appendix:B} (see Figure \ref{fig.c_T}).

\section{Thermal Skin Depth and Thermal Inertia}
\label{sec.3}

In Section \ref{sec.3}, we introduce the diurnal and orbital thermal skin depth and thermal inertia.
We also show the numerical results and compare our calculations with observations.
We note that the diurnal/orbital variations of the temperature reflect the thermophysical properties of a cometary surface shallower than the diurnal/orbital thermal skin depth.

\subsection{Diurnal thermal skin depth}
\label{sec.ds}

Based on the observations of the diurnal variation of the surface and subsurface temperatures, the thermal inertia of comet 67P/C--G was investigated by several studies \citep[e.g.,][]{Gulkis+2015,Schloerb+2015,Spohn+2015}.
The $e$-folding depth of the diurnal variation of temperature is called the diurnal thermal skin depth, $d_{\rm diu}$.

\begin{figure*}
\centering
\includegraphics[width = \textwidth]{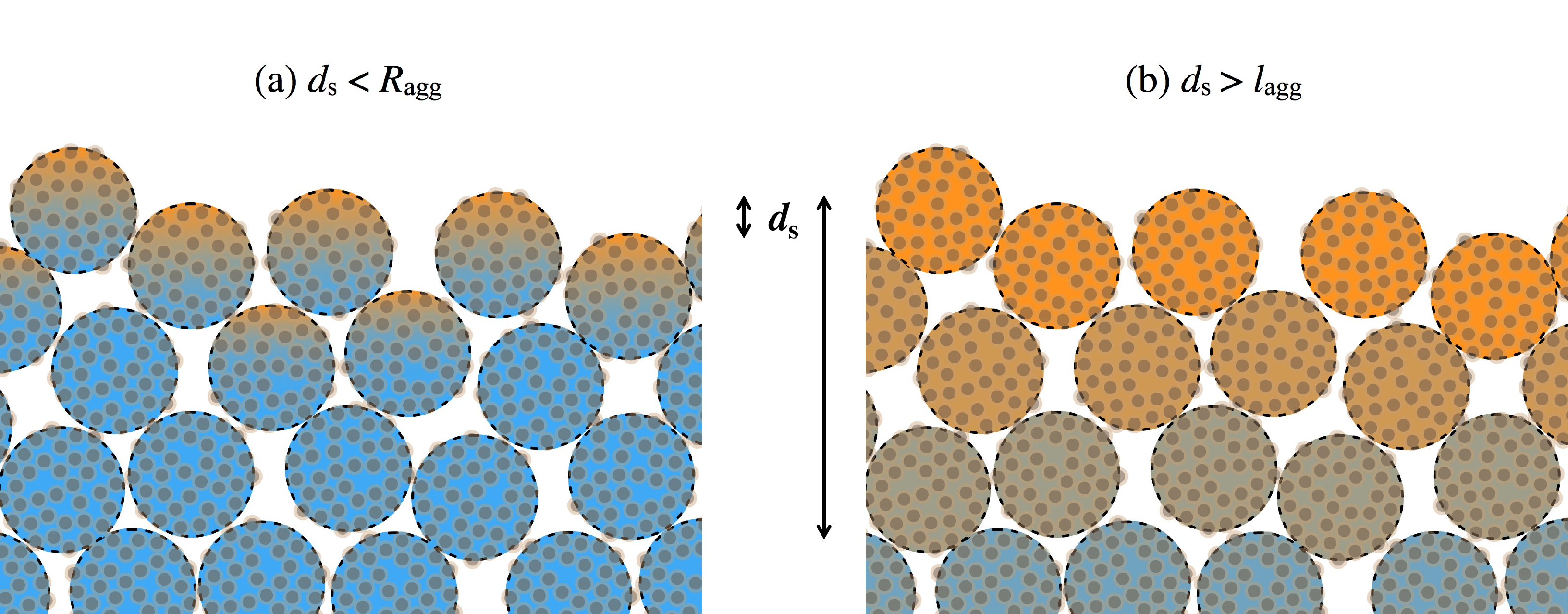}
\caption{
Schematic of the temperature structure of the cometary surface in the framework of the hierarchical aggregate model.
The color of the pebbles correspond to the temperature.
(a) If the thermal skin depth is smaller than the aggregate radius, $d_{\rm s} < R_{\rm agg}$, the observed variation of temperature should reflect the thermophysical properties of pebbles on the cometary surface.
In contrast, (b) if the thermal skin depth is larger than the typical distance among constituent aggregates, $d_{\rm s} > l_{\rm agg}$, the observed variation of temperature reflects the radiative heat transfer process within the inter-aggregate structure of hierarchical aggregates \citep{Blum+2017}.
We note that the thermal skin depth depends on the timescale of temperature variation and the diurnal and orbital thermal skin depths, $d_{\rm diu}$ and $d_{\rm orb}$, are orders of magnitude different.
}
\label{fig4}
\end{figure*}

We assert that the physical mechanism that controls thermal inertia depends on whether the aggregate size is larger or smaller than the thermal skin depth.
This is because the variation of temperature reflects the thermophysical properties of the surface region that is shallower than the thermal skin depth.
If the diurnal thermal skin depth is smaller than the aggregate radius, $d_{\rm diu} < R_{\rm agg}$, the observed diurnal variation of temperature should reflect the thermophysical properties of the pebbles on the cometary surface, as shown in Figure \ref{fig4}(a).
In this case, the diurnal thermal skin depth is given by
\begin{equation}
d_{\rm diu, agg} = \sqrt{\frac{k_{\rm agg} P_{\rm s}}{\pi c_{\rm m} \rho_{\rm m} \phi_{\rm agg}}},
\end{equation}
where $P_{\rm s} = 12.4\ {\rm h}$ is the rotation period of comet 67P/C--G \citep[e.g.,][]{Jorda+2016}.
We note that the diurnal thermal skin depth is independent of the aggregate radius, $R_{\rm agg}$, when it is given by $d_{\rm diu, agg}$.

In contrast, if the diurnal thermal skin depth is larger than the typical distance among constituent aggregates, $d_{\rm diu} > l_{\rm agg}$, the observed diurnal variation of temperature may reflect the radiative heat transfer process within the inter-aggregate structure of hierarchical aggregates \citep[e.g.,][]{Blum+2017}, as shown in Figure \ref{fig4}(b).
In this case, the diurnal thermal skin depth is given by
\begin{equation}
d_{\rm diu, hie} = \sqrt{\frac{k_{\rm hie} P_{\rm s}}{\pi c_{\rm m} \rho_{\rm m} \phi_{\rm total}}}.
\end{equation}

In this study, we assume that the diurnal thermal skin depth is given by $d_{\rm diu} = d_{\rm diu, agg}$ when the condition,
\begin{equation}
d_{\rm diu, agg} < R_{\rm agg}\ {\rm and}\ d_{\rm diu, hie} < l_{\rm agg},
\end{equation}
is satisfied.
Similarly, when the condition,
\begin{equation}
d_{\rm diu, agg} > R_{\rm agg}\ {\rm and}\ d_{\rm diu, hie} > l_{\rm agg},
\end{equation}
is satisfied, we set $d_{\rm diu} = d_{\rm diu, hie}$.
We can rewrite the equation $d_{\rm diu, hie} = l_{\rm agg}$ as
\begin{equation}\label{eq:Ragg_cri}
R_{\rm agg} = \frac{4 \sigma_{\rm SB} T^{3}}{\pi c_{\rm m} \rho_{\rm m}} \frac{1 - \phi_{\rm p}}{\phi_{\rm p}} P_{\rm s},
\end{equation}
and this equation gives the critical aggregate radius that satisfies $d_{\rm diu, hie} = l_{\rm agg}$.

\begin{figure*}
\centering
\includegraphics[width = \columnwidth]{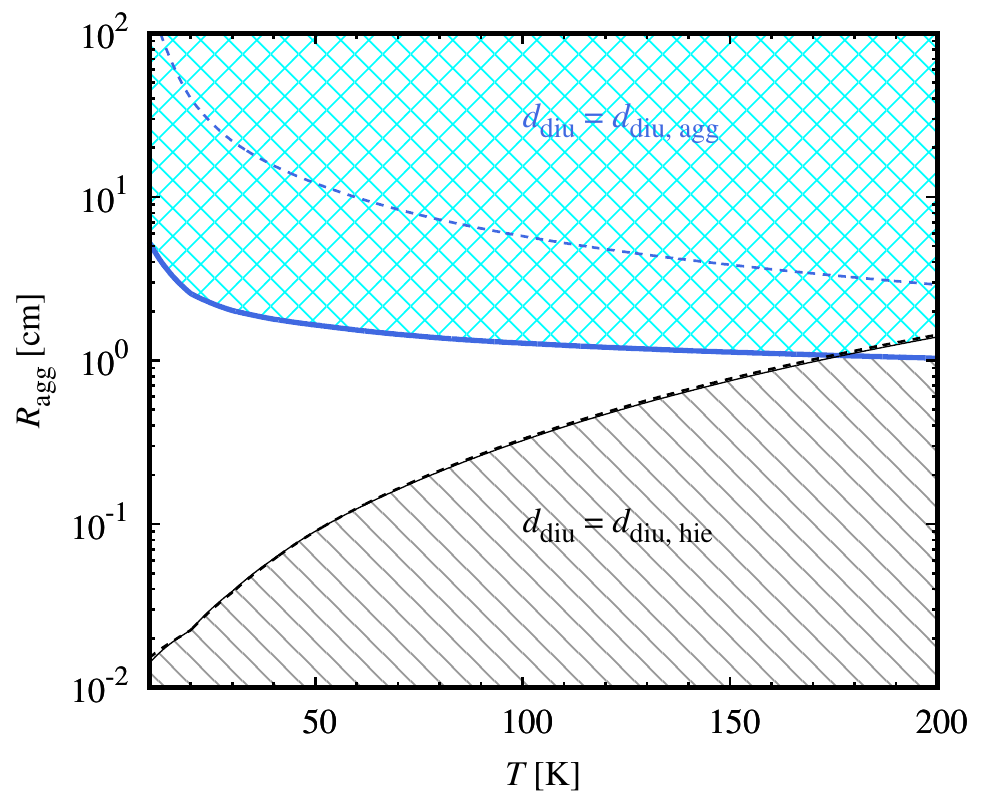}
\includegraphics[width = \columnwidth]{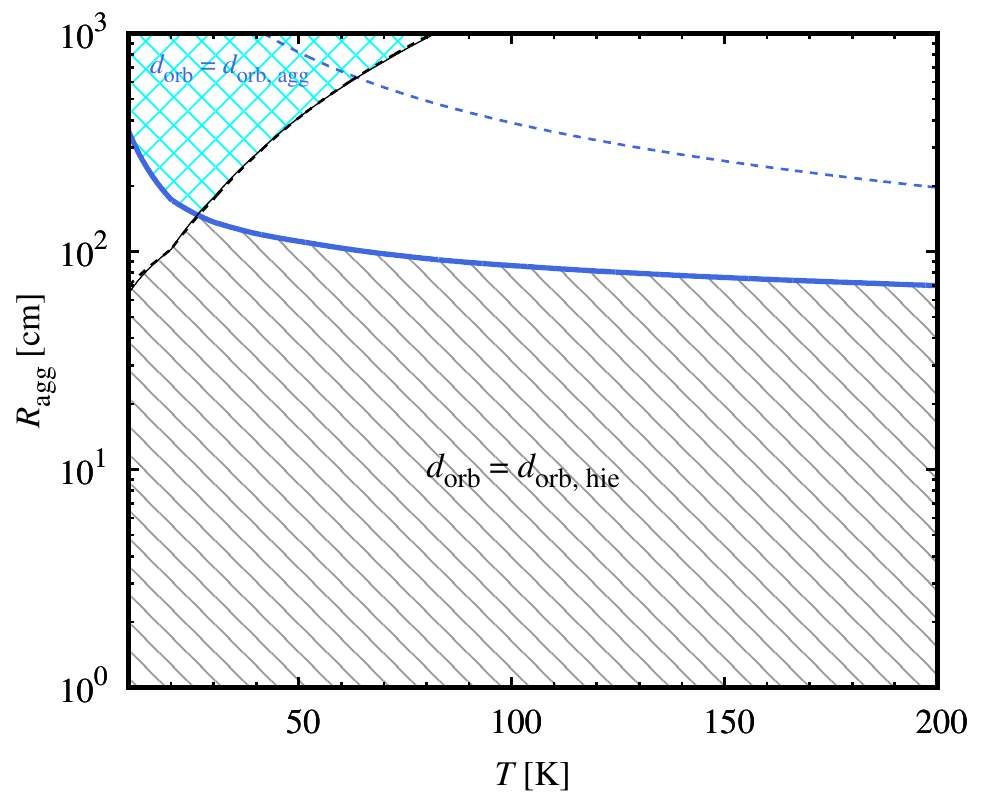}
\caption{
({\it Left panel}) The range of the aggregate radius $R_{\rm agg}$ where the diurnal thermal skin depth is given by $d_{\rm diu} = d_{\rm diu, agg}$ (cyan crosshatched region) or $d_{\rm diu} = d_{\rm diu, hie}$ (grey hatched region), for the case in which monomer grains are organic--silicate grains.
We note that a cometary surface shallower than several centimetres may be covered by the pebbles that constitute the organic--silicate grains (solid lines), based on the far ultraviolet spectrum of the cometary surface \citep{Stern+2015a}.
The blue lines represent the aggregate radius that satisfy $d_{\rm diu, agg} = R_{\rm agg}$ whereas the black lines are the solution of $d_{\rm diu, hie} = l_{\rm agg}$ (Eq. \ref{eq:Ragg_cri}).
The solid lines represent the case of organic--silicate grains, and the dashed lines are for the case of ice--organic--silicate grains.
({\it Right panel}) The range of the aggregate radius $R_{\rm agg}$ for which the orbital thermal skin depth is given by $d_{\rm orb} = d_{\rm orb, agg}$ (cyan crosshatched region) or $d_{\rm orb} = d_{\rm orb, hie}$ (grey hatched region), for the case in which monomer grains are organic--silicate grains.
The blue lines indicate the aggregate radius that satisfies $d_{\rm orb, agg} = R_{\rm agg}$, and the black lines are the solution of $d_{\rm orb, hie} = l_{\rm agg}$.
The solid lines represent the case of organic--silicate grains, whereas the dashed lines represent the case of ice--organic--silicate grains.
}
\label{fig5}
\end{figure*}

Figure \ref{fig5} shows the range of the aggregate radius $R_{\rm agg}$ where the diurnal thermal skin depth is given by $d_{\rm diu} = d_{\rm diu, agg}$ (cyan crosshatched region) or $d_{\rm diu} = d_{\rm diu, hie}$ (grey hatched region), for the case in  which monomer grains are organic--silicate grains.
The diurnal thermal skin depth is also shown in Figure \ref{fig.app.diu} (see Appendix \ref{app.ds}).
Since $k_{\rm agg}$, $k_{\rm hie}$, and $c_{\rm m}$ depend on the temperature, the diurnal thermal skin depth is dependent on the temperature.
The blue lines represent the aggregate radius that satisfies $d_{\rm diu, agg} = R_{\rm agg}$, and the black lines are the solution of $d_{\rm diu, hie} = l_{\rm agg}$.
The solid lines are associated with the case of organic--silicate grains, and the dashed lines represent the case of ice--organic--silicate grains.
For the case of organic--silicate monomers, we found that the diurnal thermal skin depth is given by $d_{\rm diu} = d_{\rm diu, agg}$ when the aggregate radius is
\begin{equation}
R_{\rm agg} \gtrsim 1\ {\rm cm},
\end{equation}
and for the case of ice--organic--silicate monomers, the critical aggregate radius is between a few centimetres and decimetres.
The large critical radius for ice--organic--silicate monomers is attributed to the high thermal conductivity of ${\rm H}_{2}{\rm O}$ ice, which is orders of magnitudes higher than that of organics.
We also assumed that ${\rm H}_{2}{\rm O}$ ice is crystalline if it exists.
Based on the Rome model for the thermal evolution of cometary nuclei \citep[e.g.,][]{Capria+2017}, at the uppermost tens of centimetres, ${\rm H}_{2}{\rm O}$ ice is crystallized and/or evaporated by the illumination history of comet 67P/C--G.

We acknowledge that for the case of (i) $d_{\rm diu, agg} < R_{\rm agg}\ {\rm and}\ d_{\rm diu, hie} > l_{\rm agg}$, or (ii) $d_{\rm diu, agg} > R_{\rm agg}\ {\rm and}\ d_{\rm diu, hie} < l_{\rm agg}$, we cannot determine the diurnal thermal skin depth at present (white regions in Figure \ref{fig5}).
It is  necessary to perform accurate numerical simulations on heat transfer within hierarchical aggregates using a discrete media approach in future research.

\subsection{Diurnal thermal inertia}

The diurnal temperature variation is inversely proportional to the diurnal thermal inertia, $I_{\rm diu}$.
Herein, we consider the diurnal thermal inertia.
When the condition for $d_{\rm diu} = d_{\rm diu, agg}$ is satisfied, the diurnal temperature variation reflects the thermal inertia of the constituent aggregates:
\begin{equation}
I_{\rm agg} = \sqrt{k_{\rm agg} c_{\rm m} \rho_{\rm m} \phi_{\rm agg}}.
\end{equation}

In contrast, when the condition for $d_{\rm diu} = d_{\rm diu, hie}$ is satisfied, the diurnal temperature variation is determined based on radiative transfer within the inter-aggregate structure \citep[e.g.,][]{Blum+2017}.
In this case, the thermal inertia of hierarchical aggregates is given by
\begin{equation}
I_{\rm hie} = \sqrt{k_{\rm hie} c_{\rm m} \rho_{\rm m} \phi_{\rm total}}.
\label{eqIhie}
\end{equation}

\begin{figure*}
\centering
\includegraphics[width = \columnwidth]{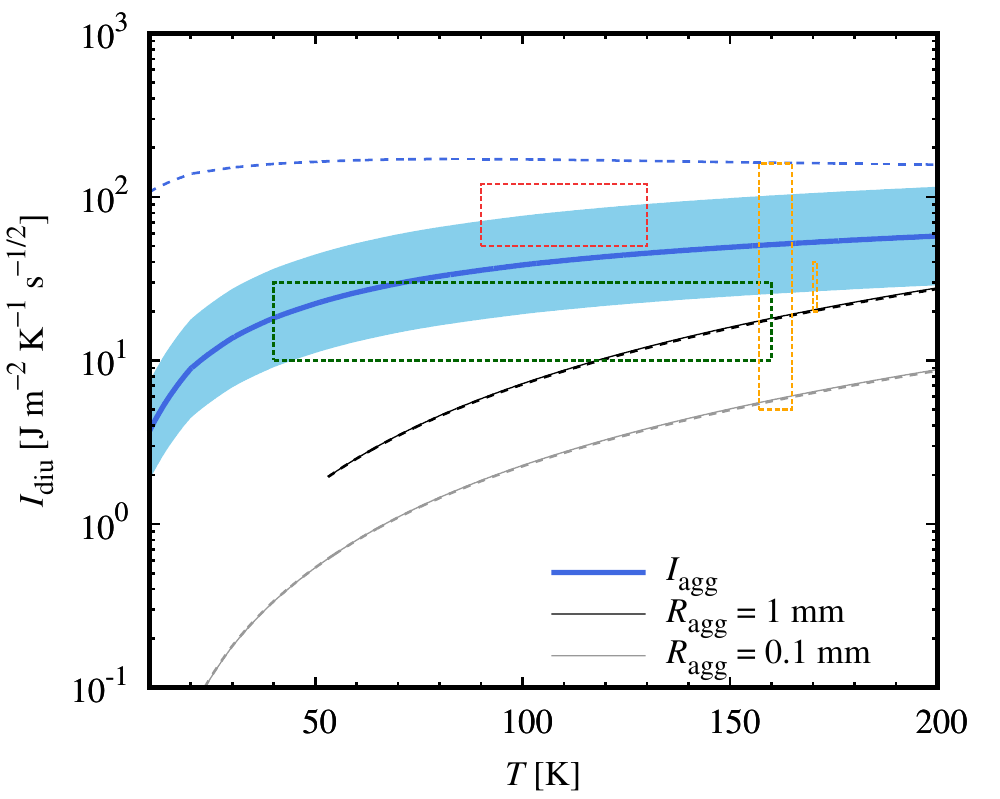}
\includegraphics[width = \columnwidth]{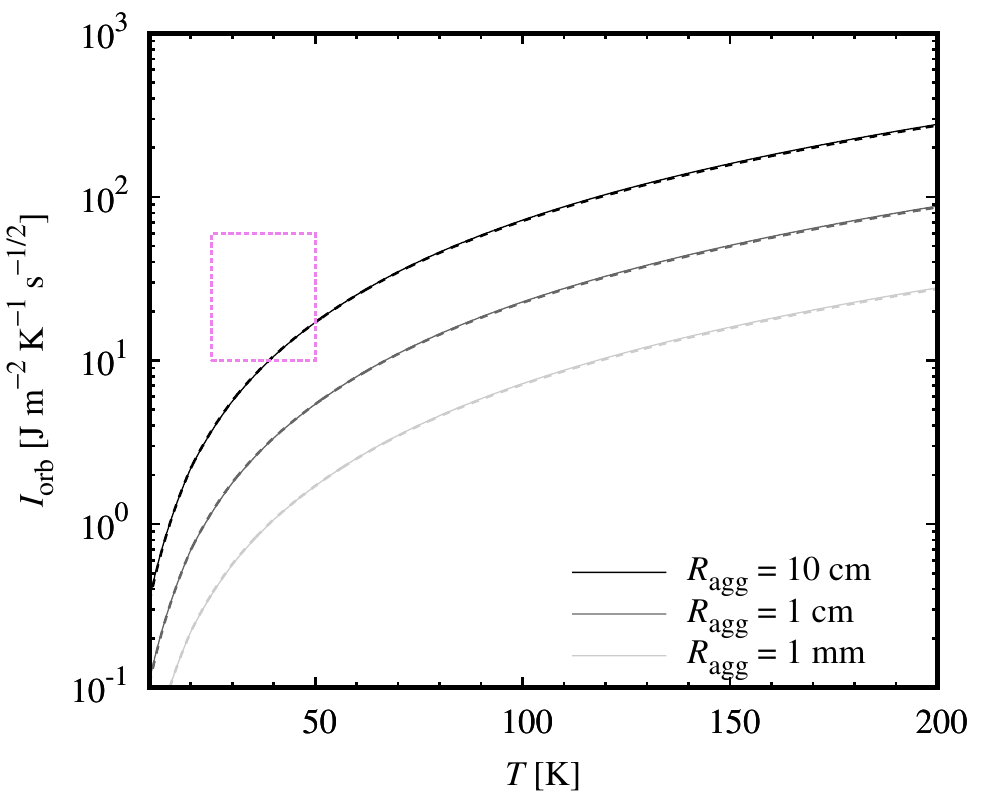}
\caption{
({\it Left panel}) Diurnal thermal inertia, $I_{\rm diu}$, as a function of temperature.
The blue lines represent the thermal inertia of the constituent aggregates, $I_{\rm agg}$.
The black lines represent the thermal inertia of hierarchical aggregates, $I_{\rm hie}$, for the case of $R_{\rm agg} = 1\ {\rm mm}$, and the grey lines represent $I_{\rm hie}$ for the case of $R_{\rm agg} = 0.1\ {\rm mm}$.
The solid lines represent the case for organic--silicate grains, and the dashed lines represent the case for ice--organic--silicate grains.
We note that a cometary surface shallower than several centimetres may be covered by pebbles that constitute organic--silicate grains (solid lines), based on the far-ultraviolet spectrum of this surface \citep{Stern+2015a}.
The blue shaded region represents the range of possible values of $I_{\rm agg}$, considering local variations of thermophysical properties, i.e., between $0.5 I_{\rm agg}$ and $2 I_{\rm agg}$
(for example, the material thermal conductivity of organics could vary by a factor of four; see Figure \ref{fig.K2017}).
The dashed boxes represent the estimated value of the diurnal thermal inertia based on observations.
The orange, green, and red boxes are attributable to the research of \citet{Marshall+2018}, \citet{Schloerb+2015}, and \citet{Spohn+2015}, respectively (see Section \ref{sec.obs.diu}).
({\it Right panel}) Orbital thermal inertia, $I_{\rm orb}$, as a function of temperature.
The colors of the lines represent the aggregate radius ($R_{\rm agg} = 10\ {\rm cm}$, $1\ {\rm cm}$, and $1\ {\rm mm}$).
In this case, the orbital thermal inertia is given by $I_{\rm orb} = I_{\rm hie}$ because $d_{\rm orb} = d_{\rm orb, hie}$ when $R_{\rm agg} \lesssim 1\ {\rm m}$.
The solid lines represent the case of organic--silicate grains, whereas the dashed lines are for the case of ice--organic--silicate grains.
The violet dashed box reprsents the estimated value of the orbital thermal inertia in the polar night regions based on observations \citep{Choukroun+2015}.
}
\label{fig6}
\end{figure*}

In this study, we set $I_{\rm diu} = I_{\rm agg}$ ($I_{\rm diu} = I_{\rm hie}$) when $d_{\rm diu} = d_{\rm diu, agg}$ ($d_{\rm diu} = d_{\rm diu, hie}$).
Figure \ref{fig6} shows the diurnal thermal inertia as a function of temperature.
The blue lines represent the thermal inertia of the constituent aggregates, $I_{\rm agg}$.
The black lines represent the thermal inertia of hierarchical aggregates, $I_{\rm hie}$, for the case of $R_{\rm agg} = 1\ {\rm mm}$, whereas the grey lines represent $I_{\rm hie}$ for the case of $R_{\rm agg} = 0.1\ {\rm mm}$.
The solid lines are for the case of organic--silicate grains, whereas the dashed lines are for the case of ice--organic--silicate grains.
We found that (i) $I_{\rm agg}$ is consistent with the observations for the case of organic--silicate grains, and (ii) when the aggregate radius is larger than $\sim 1\ {\rm mm}$, the observed diurnal thermal inertia is also consistent with $I_{\rm hie}$ in our calculations for both organic--silicate and ice--organic--silicate grains.

We note that, based on the far-ultraviolet spectrum, there is no evidence of ${\rm H}_{2}{\rm O}$ ice absorption on the cometary surface \citep{Stern+2015a}.
The observed spectrum is more consistent with the idea that the cometary surface is covered with pebbles made of organic--silicate grains (solid lines in Figures \ref{fig5} and \ref{fig6}).

It is also worth noting that the physical properties probed by the diurnal thermal inertia depend on whether $R_{\rm agg} \gtrsim 1\ {\rm cm}$.
For the case of $R_{\rm agg} \gtrsim 1\ {\rm cm}$, the diurnal thermal inertia reflects the thermal conductivity of the constituent aggregates, $k_{\rm agg}$, and $k_{\rm agg}$ is mainly dependent on the material thermal conductivity of the outermost layer of core--mantle monomers.
\footnote{
We note that the dependence of $I_{\rm agg}$ on the monomer radius $R$ is  exceedingly weak: $I_{\rm agg} \propto R^{- 1/6}$.
The dependence on the surface energy is also weak: $I_{\rm agg} \propto \gamma^{1/6}$ (see Equation \ref{eqac}).
}
In contrast, for the case of $R_{\rm agg} \lesssim 1\ {\rm mm}$, the diurnal thermal inertia reflects the thermal conductivity of hierarchical aggregates, $k_{\rm hie}$, and $k_{\rm hie}$ is proportional to the cube of the aggregate radius: $k_{\rm hie} \propto {R_{\rm agg}}^{3}$.
Therefore, we can probe $R_{\rm agg}$ using the diurnal thermal inertia.

\subsection{Orbital thermal skin depth and orbital thermal inertia}

Although most thermal observations focus on diurnal temperature variations, \citet{Choukroun+2015} investigated the orbital variation of the temperature at the southern polar regions of comet 67P/C--G.
Herein, we discuss the orbital thermal skin depth and orbital thermal inertia.
We note that the diurnal/orbital thermal skin depth is proportional to the square root of the spin/orbital period.
Therefore, the orbital thermal skin depth is several orders of magnitude larger than the diurnal thermal skin depth.

Similar to the diurnal thermal skin depth and diurnal thermal inertia, we can also define the orbital thermal skin depth, $d_{\rm orb}$, and the orbital thermal inertia, $I_{\rm orb}$.
If the orbital thermal skin depth is smaller than the aggregate radius, $d_{\rm orb} < R_{\rm agg}$, it is given by
\begin{equation}
d_{\rm orb, agg} = \sqrt{\frac{k_{\rm agg} P_{\rm o}}{\pi c_{\rm m} \rho_{\rm m} \phi_{\rm agg}}},
\end{equation}
where $P_{\rm o} = 6.45\ {\rm yr}$ is the orbital period of comet 67P/C--G (JPL Small-Body Database\footnote{
https://ssd.jpl.nasa.gov/sbdb.cgi?sstr=67P
}).
In contrast, if the orbital thermal skin depth is larger than the typical distance among constituent aggregates, $d_{\rm diu} > l_{\rm agg}$, it is given by
\begin{equation}
d_{\rm orb, hie} = \sqrt{\frac{k_{\rm hie} P_{\rm o}}{\pi c_{\rm m} \rho_{\rm m} \phi_{\rm total}}}.
\end{equation}
We assume that the diurnal thermal skin depth is given by $d_{\rm orb} = d_{\rm orb, agg}$ when
\begin{equation}
d_{\rm orb, agg} < R_{\rm agg}\ {\rm and}\ d_{\rm orb, hie} < l_{\rm agg},
\end{equation}
and also assume that the diurnal thermal skin depth is given by $d_{\rm orb} = d_{\rm orb, hie}$ when the condition,
\begin{equation}
d_{\rm orb, agg} > R_{\rm agg}\ {\rm and}\ d_{\rm orb, hie} > l_{\rm agg},
\end{equation}
is satisfied.
\footnote{
We also acknowledge that, for the case of (i) $d_{\rm orb, agg} < R_{\rm agg}\ {\rm and}\ d_{\rm orb, hie} > l_{\rm agg}$, or (ii) $d_{\rm orb, agg} > R_{\rm agg}\ {\rm and}\ d_{\rm orb, hie} < l_{\rm agg}$, we cannot determine the orbital thermal skin depth at present (white regions in Figure \ref{fig5}), as indicated in Section \ref{sec.ds}.
}

The orbital skin depth is controlled by thermal conductivity of hierarchical aggregates, because the orbital thermal skin depth is larger than the aggregate radius (see Appendix \ref{app.ds}).
The right panel of Figure \ref{fig5} shows the range of the aggregate radius $R_{\rm agg}$ where the diurnal thermal skin depth is given by $d_{\rm diu} = d_{\rm diu, agg}$ (cyan crosshatched region) or $d_{\rm diu} = d_{\rm diu, hie}$ (grey hatched region), for the case in which the monomer grains are organic--silicate grains.
As shown in Figure \ref{fig5}, the orbital thermal skin depth is given by $d_{\rm orb} = d_{\rm orb, hie}$ when the aggregate radius is
\begin{equation}
R_{\rm agg} \lesssim 1\ {\rm m}.
\end{equation}
We also set $I_{\rm orb} = I_{\rm hie}$ ($I_{\rm orb} = I_{\rm agg}$) when $d_{\rm orb} = d_{\rm orb, hie}$ ($d_{\rm orb} = d_{\rm orb, agg}$) similar to the diurnal thermal inertia.
The right panel of Figure \ref{fig6} shows the orbital thermal inertia as a function of temperature.
In this case, the orbital thermal inertia is given by $I_{\rm orb} = I_{\rm hie}$ because $d_{\rm orb} = d_{\rm orb, hie}$ when $R_{\rm agg} \lesssim 1\ {\rm m}$.

\subsection{Comparison with observational data}
\label{sec.obs.diu}

In this section, we compare the diurnal and orbital thermal inertias that were calculated using our model with that measured during the {\it Rosetta} mission.
We will show that the calculated diurnal thermal inertia reasonably explains the measured inertia when the aggregate size is larger than $\sim 1\ {\rm mm}$.
However, the large aggregate radius ($R_{\rm agg} \gtrsim 3\ {\rm cm}$) is required to reproduce the orbital thermal inertia reported by \citet{Choukroun+2015}.

\subsubsection{\citet{Marshall+2018}}

\citet{Marshall+2018} derived estimates for the diurnal thermal inertia in several regions on the largest lobe of the nucleus by analyzing data from the Microwave Instrument for the Rosetta Orbiter \citep[MIRO;][]{Gulkis+2007} and the Visible and InfraRed Thermal Imaging Spectrometer \citep[VIRTIS;][]{Coradini+2007}.
The MIRO radiometer measures antenna temperatures at millimetre ($1.6\ {\rm mm}$) and submillimetre wavelengths \citep[$0.5\ {\rm mm}$;][]{Gulkis+2007}.
The VIRTIS instrument consists of a high-spectral-resolution point spectrometer and two mapping channels, and \citet{Marshall+2018} used data acquired by the mapping channels, VIRTIS-M-IR \citep[$0.95$--$5.1\ {\mu}{\rm m}$;][]{Coradini+2007}.
The MIRO millimetre and submillimetre emissions originate from a depth of a few centimetres \citep[e.g.,][]{Schloerb+2015}, whereas the VIRTIS infrared spectrometer was most sensitive to the temperature of the uppermost few tens of microns \citep{Marshall+2018}.

\citet{Marshall+2018} calculated the vertical temperature structure of the surface and subsurface of the comet in response to insolation, then they obtained simulated brightness temperatures as a function of the diurnal thermal inertia.
The Aten region was observed via MIRO on September 2nd and 15th, 2014, and the Ash region on September 12th and 13th.
For the Aten region, the observed submillimetre brightness temperatures are $170\ {\rm K}$ and $171\ {\rm K}$, and the diurnal thermal inertia estimated from the brightness temperature calculations is in the range $20$--$40\ {\rm J}\ {\rm m}^{-2}\ {\rm K}^{-1}\ {\rm s}^{- 1/2}$.
For the Ash region, the observed submillimetre brightness temperatures are $157\ {\rm K}$ and $165\ {\rm K}$, and the diurnal thermal inertia estimated from the brightness temperature calculations is in the range $5$--$160\ {\rm J}\ {\rm m}^{-2}\ {\rm K}^{-1}\ {\rm s}^{- 1/2}$.

Figure \ref{fig6} shows that the diurnal thermal inertia estimated from submillimetre brightness temperatures is consistent with our calculation of $I_{\rm diu}$.
The observed $I_{\rm diu}$ for the Aten and Ash regions (orange dashed boxes) can be reproduced when (i) $I_{\rm diu} = I_{\rm agg}$ (i.e., $R_{\rm agg} \gtrsim 1\ {\rm cm}$) and the monomers are organic--silicate grains (blue shaded region), or (ii) $I_{\rm diu} = I_{\rm hie}$ and the aggregate radius is larger than $1\ {\rm mm}$ (black lines).

Based on millimetre brightness temperatures, \citet{Marshall+2018} also estimated the upper limit of the diurnal thermal inertia as $80\ {\rm J}\ {\rm m}^{-2}\ {\rm K}^{-1}\ {\rm s}^{- 1/2}$ for both the Aten and Ash regions.
In contrast, VIRTIS observations suggest a best-fitting value of $80\ {\rm J}\ {\rm m}^{-2}\ {\rm K}^{-1}\ {\rm s}^{- 1/2}$ across the observed Aten, Babi, Khepry, and Imhotep regions.
These values of $I_{\rm diu}$ is also consistent with our calculations when $I_{\rm diu} = I_{\rm agg}$ and the monomers are organic--silicate grains (blue shaded region in Figure \ref{fig6}).

\subsubsection{\citet{Schloerb+2015}}

\citet{Schloerb+2015} analyzed the observed brightness temperatures as a function of local solar time and effective latitude, which is based on the orientation of the local surface normal of a point on the surface with respect to the sun.
All MIRO observations obtained during the period September 1--30, 2014 were included in \citet{Schloerb+2015}.
The MIRO emission exhibits strong diurnal variations, which indicate that it originates from within the thermally varying layer in the upper centimetres of the surface.

A comparison of the mean MIRO brightness temperatures to the predictions of the thermal models reveals good agreement for most latitudes (from $- 20$ to $40$ degrees), for which the mean temperature is in the range $\sim 90$--$160\ {\rm K}$ \citep[see Figure 9 of][]{Schloerb+2015}.
\footnote{
\citet{Schloerb+2015} noted that the MIRO brightness temperatures at high northern latitudes are compatible with the fact that sublimation of ${\rm H}_{2}{\rm O}$ ice playing an important role in determining the temperatures of these regions, wherein based on observations of gas and dust production, ice is known to sublimate.
However, the thermal model used in \citet{Schloerb+2015} did not consider this effect and advanced thermophysical modelling is required to understand the brightness temperatures at high northern latitudes.
}
The quantitative fit of simple thermophysical models is consistent with the diurnal thermal inertia in the range $10$--$30\ {\rm J}\ {\rm m}^{-2}\ {\rm K}^{-1}\ {\rm s}^{- 1/2}$ and  the diurnal thermal skin depth is approximately $1\ {\rm cm}$.
The estimated $I_{\rm diu}$ by \citet{Schloerb+2015}, the green dashed box in Figure \ref{fig6}, can be reproduced when (i) $I_{\rm diu} = I_{\rm agg}$ (i.e., $R_{\rm agg} \gtrsim 1\ {\rm cm}$) and the monomers are organic--silicate grains (blue shaded region), or (ii) $I_{\rm diu} = I_{\rm hie}$ and the aggregate radius is larger than $1\ {\rm mm}$ (black lines).

\subsubsection{\citet{Spohn+2015}}

The Multipurpose Sensors for Surface and Sub-Surface Science \citep[MUPUS;][]{Spohn+2007} instrument package was operated on the approach to and on the surface of 67P/C--G during November 12--14, 2014.
\citet{Spohn+2015} found that the diurnal temperature at the {\it Philae} landing site, Abydos, varied between $90\ {\rm K}$ and $130\ {\rm K}$, and the local thermal inertia was $85 \pm 35\ {\rm J}\ {\rm m}^{-2}\ {\rm K}^{-1}\ {\rm s}^{- 1/2}$. 
Although the estimated thermal inertia is higher than the MIRO measurements, this could be explained by heterogeneities in the surface layer, e.g., filling factor, temperature, refractory-to-ice mass ratio, and other factors.
The observed $I_{\rm diu}$ at the Abydos site (red dashed box in Figure \ref{fig6}) can be reproduced when $I_{\rm diu} = I_{\rm agg}$ (i.e., $R_{\rm agg} \gtrsim 1\ {\rm cm}$) and the monomers are organic--silicate grains (blue shaded region).
In addition, not only organic--silicate grains, but also ice--organic--silicate monomer grains could explain the observed $I_{\rm diu}$ at the Abydos site when $I_{\rm diu} = I_{\rm agg}$ (i.e., $R_{\rm agg} \gtrsim 10\ {\rm cm}$; blue dashed line).

\subsubsection{\citet{Choukroun+2015}}

\citet{Choukroun+2015} reported on observations made with the submillimetre and millimetre continuum channels of the MIRO of the thermal emission from the southern regions of the nucleus during the period August--October 2014.
Since the southern polar regions were in darkness for five years, subsurface temperatures in the range $25$--$50\ {\rm K}$ were measured.

Based on their thermal model calculations of the nucleus near-surface temperatures conducted over the orbit of comet 67P/C--G, \citet{Choukroun+2015} revealed that the southern polar regions have a thermal inertia within the range $10$--$60\ {\rm J}\ {\rm m}^{-2}\ {\rm K}^{-1}\ {\rm s}^{- 1/2}$.
Diurnal phase effects are absent in the polar night regions, and the thermal inertia obtained by \citet{Choukroun+2015} reflects the orbital thermal inertia, $I_{\rm orb}$.

As shown in Figure \ref{fig6}, an aggregate radius of $R_{\rm agg} > 3\ {\rm cm}$ is required to explain $I_{\rm orb} > 10\ {\rm J}\ {\rm m}^{-2}\ {\rm K}^{-1}\ {\rm s}^{- 1/2}$ at $T = 50\ {\rm K}$.
\footnote{
We confirmed that, at the temperature of $T = 50\ {\rm K}$, the condition, $k_{\rm agg} > k_{\rm hie}$, is satisfied when the aggregate radius is $R_{\rm agg} < 11\ {\rm cm}$ for the case of organic--silicate grains (and $R_{\rm agg} < 5.9\ {\rm m}$ for the case of ice--organic--silicate grains; see Section \ref{sec.k.hie}).
}
Since the thermal conductivity of hierarchical aggregates associated with radiative transfer within inter-aggregate voids, $k_{\rm hie}$, is independent of the monomer composition, the aggregate radius required to explain the reported thermal inertia is insensitive to whether the monomers are organic--silicate grains or ice--organic--silicate grains.
We conclude that the pebbles on the southern polar regions should be cm- or dm-sized to reproduce the orbital thermal inertia reported by \citet{Choukroun+2015}.

\subsection{Summary of the thermal inertia calculations}

We found that the thermal inertia depends on the temperature and the timescale of the temperature variation.
Therefore, we define $I_{\rm diu}$ and $I_{\rm orb}$ for diurnal and orbital temperature variations, respectively.
The heat transfer process depends on whether the thermal skin depth is smaller than the aggregate radius or not, as shown in Figure \ref{fig4}.

Our calculations revealed that, when $1\ {\rm cm} \lesssim R_{\rm agg} \lesssim 1\ {\rm m}$, the diurnal thermal inertia is given by the thermal inertia of the pebbles, $I_{\rm agg}$, whereas the orbital thermal inertia is given by the thermal inertia of the hierarchical aggregates due to radiative transfer within inter-aggregate voids, $I_{\rm hie}$.
The value of the calculated $I_{\rm agg}$ is consistent with the observed diurnal thermal inertia in various regions \citep{Schloerb+2015,Spohn+2015,Marshall+2018}, and the observed orbital thermal inertia can be reproduced when the aggregate radius is larger than $3\ {\rm cm}$.
Therefore, hierarchical aggregates of cm- to dm-sized (i.e., $3\ {\rm cm} \lesssim R_{\rm agg} < 1\ {\rm m}$) pebbles can explain the thermal inertia of comet 67P/C--G.

\section{Discussion: other estimates of the size of the pebbles}
\label{sec.5}

Our estimate of the size of the pebbles is consistent with the constraint on the physical homogeneity of comet 67P/C--G \citep[e.g.,][]{Kofman+2015,Paetzold+2016}.
\citet{Kofman+2015} obtained Comet Nucleus Sounding Experiment by Radiowave Transmission (CONSERT) measurements of the interior of comet 67P/C--G, and they found that the interior is homogeneous on a spatial scale of $10\ {\rm m}$.
The gravity field observations also support the idea that the nucleus has a homogeneous density down to a scale of several metres \citep{Paetzold+2016}

\begin{figure}
\centering
\includegraphics[width = \columnwidth]{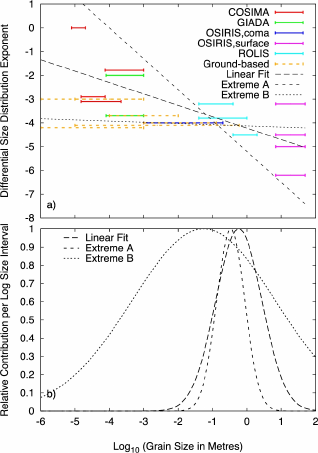}
\caption{
Size distribution of dust aggregates for comet 67P/C--G.
(a) Exponent of the size-frequency distribution function of the dust emitted from the nucleus.
Data (horizontal lines) are obtained from measurements by various instruments onboard {\it Rosetta} (solid lines) and from Earth-based observations (orange dashed lines).
The three lines represent linear fits to the data and the "Linear Fit" is the fiducial case of \citet{Blum+2017}.
(b) Derived normalized mass-frequency distributions per logarithmic size interval for the three linear approximations.
Figure taken from \citet{Blum+2017}.
}
\label{fig.Blum2017}
\end{figure}

In addition, the size-frequency distribution of dust aggregates emitted from the nucleus also support the notion that the constituent aggregates of comet 67P/C--G is cm- to dm-sized pebbles.
Figure \ref{fig.Blum2017} shows the size distribution of dust aggregates for comet 67P/C--G measured using different methods \citep[see][for details]{Blum+2017}.
It is evident that most of the mass is emitted in the form of dm-sized dust aggregates, and that there is a significant decline in the size-frequency distribution for sizes below $1\ {\rm cm}$.
\citet{Blum+2017} interpreted mm-sized dust aggregates as ``pebble fragments'' due to the ejection process.
As such, the size of the primary building blocks of the comet nucleus must be cm- or dm-sized pebbles.

\citet{Gundlach+2015} estimated the maximum radius of constituent aggregates that can be released from the cometary surface, $R_{\rm agg, max}$.
The ejected aggregates are lifted up by the gas-friction force, $F_{\rm gas}$, and $F_{\rm gas}$ must overcome the gravitational force, $F_{\rm grav}$.
The gas-friction force at the cometary surface is approximately given by $F_{\rm gas} = \pi {R_{\rm agg}}^{2} p_{\rm gas}$, where $p_{\rm gas}$ is the gas pressure at the ice sublimation interface.
The gravitational force is $F_{\rm grav} = {G M_{\rm agg} m_{\rm 67P}} / {{r_{\rm 67P}}^{2}}$, where $G$ is the gravitational constant, $r_{\rm 67P}$ is the radius of the comet, and $M_{\rm agg}$ and $m_{\rm 67P}$ are the mass of the constituent aggregates and comet 67P/C--G, respectively.
The maximum radius of constituent aggregates, $R_{\rm agg, max}$, is given by
\begin{equation}
R_{\rm agg, max} = \frac{9 p_{\rm gas}}{16 \pi G \rho_{\rm agg} \rho_{\rm 67P} r_{\rm 67P}}.
\label{eq.G15.R}
\end{equation}
\citet{Gundlach+2015} also derived a simple analytic formula for the gas pressure at the ice sublimation interface:
\begin{equation}
p_{\rm gas} = {\left( 1 - A \right)} \frac{S_{\odot} {\left( D_{\rm hel} / 1\ {\rm au} \right)}^{-2}}{\Lambda} \sqrt{\frac{2 \pi k_{\rm B} T_{\rm ice}}{m_{\rm g}}},
\label{eq.G15.p}
\end{equation}
where $A$ is the Bond albedo of the cometary surface, $S_{\odot} = 1.37 \times 10^{3}\ {\rm W}\ {\rm m}^{-2}$ is the solar constant, $D_{\rm hel}$ is the heliocentric distance of the comet, $\Lambda$ is the latent heat of sublimation, and $T_{\rm ice}$ is the temperature of the evaporating volatiles.

Using Equations (\ref{eq.G15.R}) and (\ref{eq.G15.p}), \citet{Gundlach+2015} revealed that the maximum radius of constituent aggregates is
\begin{equation}
R_{\rm agg, max} \sim 10^{2}\ {\left( \frac{D_{\rm hel}}{1\ {\rm au}} \right)}^{-2}\ {\rm cm},
\end{equation}
for both ${\rm H}_{2}{\rm O}$ and ${\rm C}{\rm O}_{2}$ activities \citep[see Figure 3 of][]{Gundlach+2015}.
The estimated value is also consistent with the size-frequency distribution of the dust aggregates emitted from the nucleus and the constraint from thermal inertias.

\section{Summary}
\label{sec.6}

We have investigated whether the hierarchical aggregate model \citep[e.g.,][]{Skorov+2012} can reproduce the observed thermal inertias of comet 67P/C--G.
Based on numerical simulations of heat transfer within dust aggregates \citep[e.g.,][]{Arakawa+2019a}, we have constructed a thermal inertia model for hierarchical aggregates.
Our findings are summarized as follows.

\begin{enumerate}
\item{
We proposed that the heat transfer process depends on whether the thermal skin depth is smaller than the aggregate radius (see Figure \ref{fig4}).
Since the diurnal and orbital thermal skin depths are different by orders of magnitude, the diurnal and orbital thermal inertias can also be controlled by different processes. 
}
\item{
Our calculations revealed that when $R_{\rm agg} \gtrsim 1\ {\rm cm}$, the diurnal thermal inertia is given by the thermal inertia of the pebbles, $I_{\rm agg}$, and if $R_{\rm agg} \lesssim 1\ {\rm mm}$, the diurnal thermal inertia is given by the thermal inertia of hierarchical aggregates due to radiative transfer within inter-aggregate voids, $I_{\rm hie}$.
In contrast, the orbital thermal inertia is always given by $I_{\rm hie}$ because the size of the pebbles is smaller than $1\ {\rm m}$.
}
\item{
The diurnal thermal inertia calculated using our model reasonably explains the thermal inertia measured during the {\it Rosetta} mission in various regions \citep{Schloerb+2015,Spohn+2015,Marshall+2018} when the aggregate radius is larger than $1\ {\rm mm}$.
}
\item{
However, the observed orbital thermal inertia in the polar night regions \citep{Choukroun+2015} could be reproduced only when the aggregate radius is larger than $3\ {\rm cm}$.
}
\end{enumerate}

Therefore, we conclude that a hierarchical aggregate of cm- to dm-sized (i.e., $3\ {\rm cm} \lesssim R_{\rm agg} < 1\ {\rm m}$) pebbles may be suitable to explain the thermal inertias of comet 67P/C--G.
We note that our estimate of the size of the pebbles is consistent with (1) the constraint on the physical homogeneity of comet 67P/C--G \citep[e.g.,][]{Kofman+2015,Paetzold+2016}, (2) the size-frequency distribution of dust aggregates emitted from the nucleus \citep[][and references therein]{Blum+2017}, and (3) the maximum radius of the constituent aggregates that can be released from the cometary surface \citep[e.g.,][]{Gundlach+2015}.

\section*{Acknowledgements}

We sincerely thank Taishi Nakamoto, Shigeru Ida, Satoshi Okuzumi, Kenji Ohta, and Hidenori Genda for insightful comments and careful reading of the draft.
We would like to thank Xinting Yu for helpful comments on the spectral analogues for cometary organics, and Eiichiro Kokubo, Hideko Nomura, and Yasuhito Sekine for useful comments and discussions.
S.A.~is supported by JSPS KAKENHI Grants Nos.~JP17J06861 and JP20J00598.
K.O.~is supported by JSPS KAKENHI Grants Nos.~JP18J14557 and JP19K03926.

\section*{Data availability}

The data underlying this article will be shared on reasonable request to the corresponding author.






\bibliographystyle{mnras}

\bibliography{references}



\appendix

\section{Physical properties used in the thermophysical models}
\label{appendix:A}

In Appendix \ref{appendix:A}, we summarize the physical properties used in this study.
Tables \ref{table1} and \ref{table2} represent overviews of the model parameters used in the thermophysical calculations.
The material density and mass fraction values are discussed in Section \ref{sec.density}.
The other mechanical properties of the materials (Poisson's ratio, surface energy, and Young's modulus) are taken from the literature.

\begin{table*}
\centering
\caption{
Materials properties used in the thermophysical models.
}
\begin{tabular}{  l  c  c  l  } \hline
Properties                                                       & Symbol                & Value                          &  Reference \\ \hline \hline
Material density of silicate                                     & $\rho_{\rm sil}$      & $3500\ {\rm kg}\ {\rm m}^{-3}$ &  Section \ref{sec.density} \\ 
Material density of organics                                     & $\rho_{\rm org}$      & $1500\ {\rm kg}\ {\rm m}^{-3}$ &  Section \ref{sec.density} \\ 
Material density of ${\rm H}_{2}{\rm O}$ ice                     & $\rho_{\rm ice}$      & $920\ {\rm kg}\ {\rm m}^{-3}$  &  --- \\ 
Poisson's ratio of organics (tholin)                             & $\nu_{\rm org}$       & $0.3$                          &  \citet{Yu+2017} \\ 
Poisson's ratio of ${\rm H}_{2}{\rm O}$ ice                      & $\nu_{\rm ice}$       & $0.25$                         &  \citet{Dominik+1997} \\ 
Surface energy of organics (tholin)                              & $\gamma_{\rm org}$    & $70.9\ {\rm mJ}\ {\rm m}^{-2}$ &  \citet{Yu+2017} \\ 
Surface energy of ${\rm H}_{2}{\rm O}$ ice                       & $\gamma_{\rm ice}$    & $20\ {\rm mJ}\ {\rm m}^{-2}$   &  \citet{Gundlach+2018} \\ 
Young's modulus of organics (tholin)                             & $E_{\rm org}$         & $3.0\ {\rm GPa}$               &  \citet{Yu+2017} \\ 
Young's modulus of ${\rm H}_{2}{\rm O}$ ice                      & $E_{\rm ice}$         & $7.0\ {\rm GPa}$               &  \citet{Dominik+1997} \\ \hline
Material thermal conductivity of organics (PMMA)                 & $k_{\rm org}$         & Fig.\ \ref{fig.k_T}            &  \citet{Choy1977} \\ 
Material thermal conductivity of ${\rm H}_{2}{\rm O}$ ice        & $k_{\rm ice}$         & Fig.\ \ref{fig.k_T}            &  \citet{Klinger1975} \\ 
Specific heat capacity of silicate (${\rm Si}{\rm O}_{2}$ glass) & $c_{\rm sil}$         & Fig.\ \ref{fig.c_T}            &  \citet{Lord+1957} \\ 
Specific heat capacity of organics (PMMA)                        & $c_{\rm org}$         & Fig.\ \ref{fig.c_T}            &  \citet{Gaur+1982} \\ 
Specific heat capacity of ${\rm H}_{2}{\rm O}$ ice               & $c_{\rm ice}$         & Fig.\ \ref{fig.c_T}            &  \citet{Shulman2004} \\ \hline
\end{tabular}
\label{table1}
\end{table*}

We set the mechanical properties of the organics to be same as that of the Titan aerosol analogue called tholin \citep{Yu+2017}.
Tholin has also been used as organic analogues for comets and Kuiper belt objects \citep[e.g.,][]{Lamy+87,Ishiguro+07,Morea+09}, and its optical property may qualitatively explain the reflectance spectrum of comet 67P/C--G \citep[e.g.,][]{Stern+2015a,Capaccioni+15}.
Then we used the mechanical and optical properties of tholin as an analogue of the cometary organics.
We also note that the mechanical properties of tholin are within the range of typical organic materials; the surface energy of typical organics is of the order of $10$--$100\ {\rm mJ}\ {\rm m}^{-2}$ \citep[e.g.,][]{Fowkes1964}, Young's modulus is of the order of $1$--$10\ {\rm GPa}$ \citep[e.g.,][]{Yu+2018}, and the Poisson's ratio is in the range of $0.3$--$0.35$ \citep[e.g.,][]{Krijt+2013}.

\begin{table*}
\centering
\caption{
Physical properties of monomer grains, aggregates, and comet 67P/C--G used in the thermophysical models.
}
\begin{tabular}{  l  c  c  l  } \hline
Properties                                                          & Symbol                                        & Value                &  Reference \\ \hline \hline
Radius of silicate cores                                            & $R_{\rm sil}$                                 & $0.5\ {\mu}{\rm m}$  &  --- \\ 
Mass fraction of organics and silicate (org--sil grains)            & $f_{\rm org}$, $f_{\rm sil}$                  & $1/3$, $2/3$         &  Section \ref{sec.density} \\
Mass fraction of ice, organics, and silicate (ice--org--sil grains) & $f_{\rm ice}$, $f_{\rm org}$, $f_{\rm sil}$   & $0.1$, $0.3$, $0.6$  &  Section \ref{sec.density} \\ \hline
Filling factor of the constituent aggregates                 & $\phi_{\rm agg}$      & $0.4$            &  \citet{Guettler+2010} \\
Filling factor of the aggregate packing structure         & $\phi_{\rm p}$        & $0.64$           &  \citet{Berryman1983} \\ \hline
Spin period of comet 67P/C--G                             & $P_{\rm s}$           & $12.4\ {\rm h}$  &  \citet{Jorda+2016} \\ 
Orbital period of comet 67P/C--G                          & $P_{\rm o}$           & $6.45\ {\rm yr}$ &  JPL Small-Body Database \\
\hline
\end{tabular}
\label{table2}
\end{table*}

\section{Material thermal conductivity and specific heat capacity}
\label{appendix:B}

\begin{figure}
\centering
\includegraphics[width = \columnwidth]{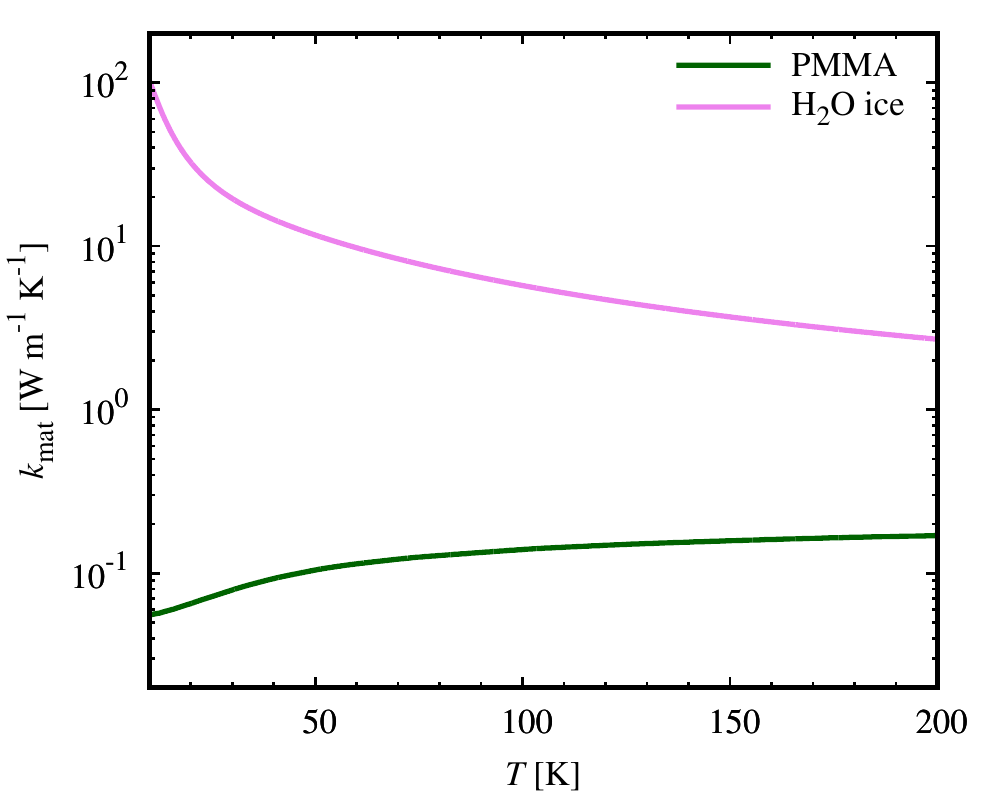}
\caption{
The material thermal conductivities of PMMA ($k_{\rm org}$, green) and ${\rm H}_{2}{\rm O}$ ice ($k_{\rm ice}$, violet).
}
\label{fig.k_T}
\end{figure}

In Appendix \ref{appendix:B}, we show the temperature dependence of material thermal conductivity and specific heat capacity.
Figure \ref{fig.k_T} shows the temperature dependence of the material thermal conductivities.
We set the material thermal conductivity of the organics as that of poly(methyl methacrylate), hereinafter referred to as PMMA.

\begin{figure}
\centering
\includegraphics[width = 0.8\columnwidth]{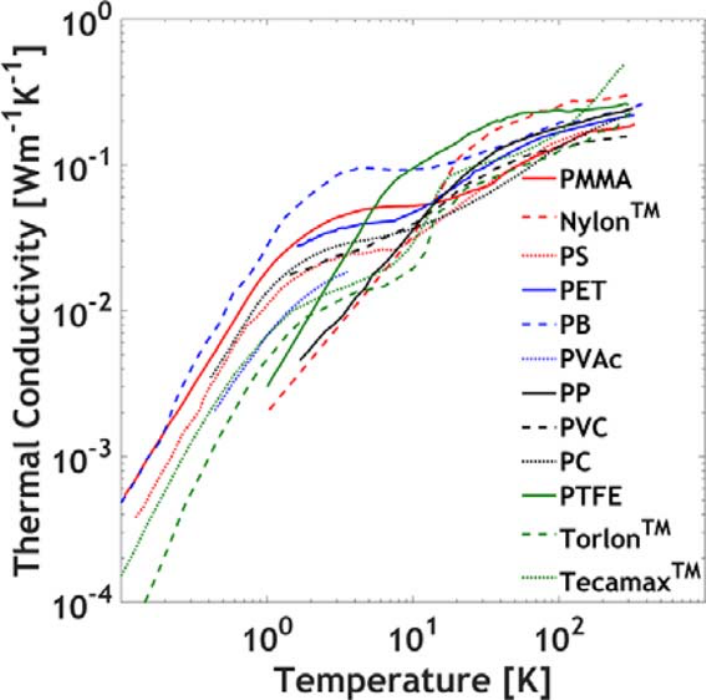}
\caption{
The material thermal conductivities for 12 different organic polymers.
Figure taken from \citet{Kommandur+2017}.
}
\label{fig.K2017}
\end{figure}

We note that $k_{\rm org}$ reliably represents the typical value of the material thermal conductivity of organics.
Figure \ref{fig.K2017} shows the material thermal conductivities of 12 different organic polymers.
In the temperature range of $T > 10\ {\rm K}$, the difference between the material thermal conductivity is within a factor of four, and $k_{\rm org} \sim 0.1\ {\rm W}\ {\rm m}^{-1}\ {\rm K}^{-1}$ when the temperature is $T \sim 100\ {\rm K}$.

\begin{figure}
\centering
\includegraphics[width = \columnwidth]{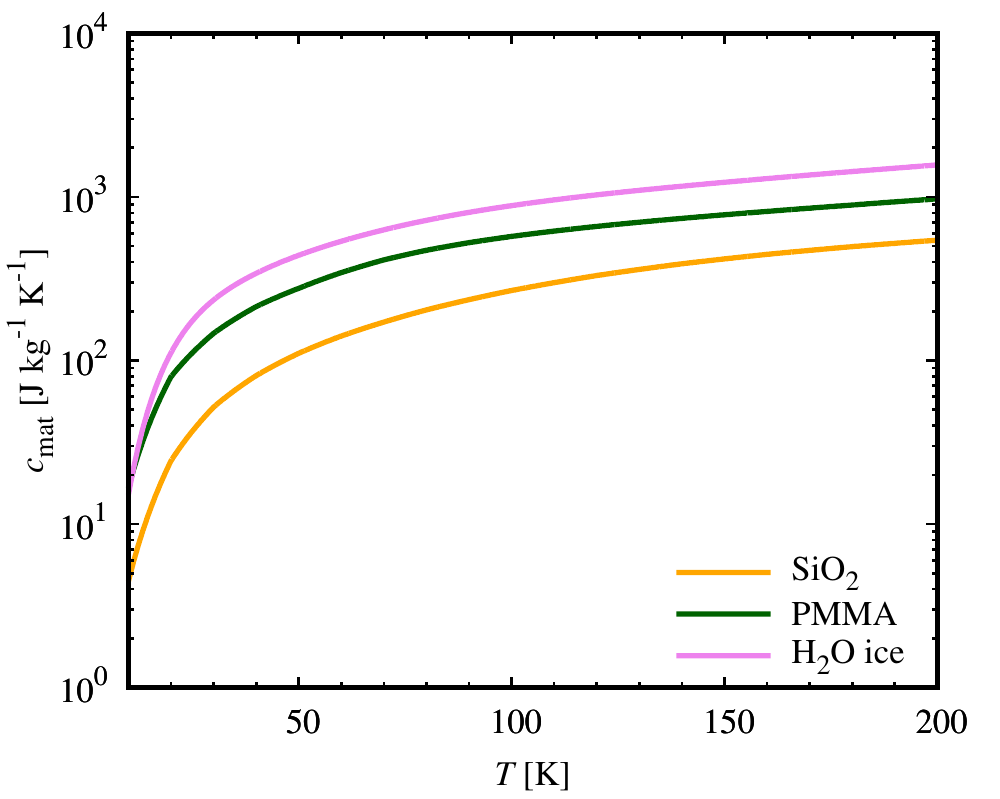}
\caption{
Specific heat capacities of ${\rm Si}{\rm O}_{2}$ glass ($c_{\rm sil}$, orange), PMMA ($c_{\rm org}$, green), and ${\rm H}_{2}{\rm O}$ ice ($c_{\rm ice}$, violet).
}
\label{fig.c_T}
\end{figure}

Figure \ref{fig.c_T} shows the temperature dependence of the specific heat capacities.
Since the specific heat capacities are approximately proportional to $T$ and $k_{\rm agg}$ is proportional to the cube of $T$, the thermal inertia of the hierarchical aggregates due to radiative transfer within the inter-aggregate structure, $I_{\rm hie}$, is approximately proportional to the square of $T$.

\section{Thermal conductivity due to radiative transfer inside constituent aggregates}
\label{app.rad}

In Appendix \ref{app.rad}, for the constituent aggregates (i.e., pebbles), we demonstrate that the thermal conductivity based on radiative transfer is negligible compared to that based on the solid network.
The thermal conductivity of the pebbles, $k_{\rm agg}$, is given by the sum of two terms:
\begin{equation}
k_{\rm agg} = k_{\rm sol} + k_{\rm rad},
\end{equation}
where $k_{\rm sol}$ is the thermal conductivity through the solid network and $k_{\rm rad}$ is the thermal conductivity due to radiative transfer within the constituent aggregate.

The thermal conductivity due to radiative transfer, $k_{\rm rad}$, is given by \citep[e.g.,][]{Merrill1969}
\begin{equation}
k_{\rm rad} = \frac{16}{3} \sigma_{\rm SB} T^{3} l_{\rm mfp, agg},
\end{equation}
where $l_{\rm mfp, agg}$ is the mean free path of photons within the constituent aggregate.
The mean free path of photons is given by \citep[e.g.,][]{Arakawa+2017}
\begin{equation}
l_{\rm mfp, agg} = \frac{1}{\kappa_{\rm R} \rho_{\rm m} \phi_{\rm agg}}.
\end{equation}
The Rosseland mean opacity, $\kappa_{\rm R}$, is defined as
\begin{equation}
\frac{1}{\kappa_{\rm R}} = \frac{\int {\rm d}\nu\ {\kappa_{\rm eff}}^{-1} {\left( {\partial B_{\nu}} / {\partial T} \right)}}{\int {\rm d}\nu\ {\left( {\partial B_{\nu}} / {\partial T} \right)}},
\label{eq:kappa_R}
\end{equation}
where $\kappa_{\rm eff}$ is the effective absorption opacity and $B_{\nu}$ is the Plank function.
\citet{Rybicki+1979} introduced the effective absorption opacity, $\kappa_{\rm eff}$, defined as $\kappa_{\rm eff} = \sqrt{\kappa_{\rm abs} {\left( \kappa_{\rm abs} + \kappa_{\rm sca} \right)}}$, where $\kappa_{\rm abs}$ is the absorption opacity and $\kappa_{\rm sca}$ is the scattering opacity.

However, forward scattering does not change the direction of the incident light, and it is effectively not scattering. 
Therefore, we use the ``effective scattering opacity'', $\kappa_{\rm sca}^{\rm eff}$, instead of $\kappa_{\rm sca}$, which is given by $\kappa_{\rm sca}^{\rm eff} = {\left( 1 - g \right)} \kappa_{\rm sca}$, where $g$ is the asymmetry parameter \citep[see][]{Ueda+2020}.
We then define $\kappa_{\rm eff}$ as
\begin{equation}
\kappa_{\rm eff} = \sqrt{\kappa_{\rm abs} {\left( \kappa_{\rm abs} + \kappa_{\rm sca}^{\rm eff} \right)}}.
\end{equation}

We calculate $\kappa_{\rm abs}$ and $\kappa_{\rm sca}$ of the spherical monomers using Mie theory \citep{Bohren+1983} using the open source code LX-MIE \citep{Kitzmann+2018}.
The refractive index of monomer grains with a core--mantle structure is calculated based on the effective medium theory with the Bruggeman mixing rule \citep{Bruggeman1935}, which is given by
\begin{equation}
\label{eq:bruggeman}
\sum_{i} \chi_{i} \frac{\epsilon_{i} - \epsilon_{\rm eff}}{\epsilon_{i} + 2 \epsilon_{\rm eff}} = 0,
\end{equation}
where $\epsilon_{i}$ is the dielectric function of the material species $i$ (silicate, organics, and ice), and $\epsilon_{\rm eff}$ is the effective dielectric function.
We note that the dielectric function satisfies the relation $\epsilon = {(n + i k)}^{2}$, where $n$ and $k$ are the real and imaginary parts of the refractive index \citep{Bohren+1983}.
To obtain the effective refractive index of the monomers, we use the refractive index of the so-called astronomical silicate \citep{Draine2003}, ${\rm H}_{2}{\rm O}$ ice \citep{Warren+08}, and Titan-tholin \citep{Khare+1984}.
We then calculate the Rosseland mean opacity by integrating Equation \eqref{eq:kappa_R} from $\lambda = 0.1\ \mu{\rm m}$ to $1000\ \mu{\rm m}$.

\begin{figure}
\centering
\includegraphics[width = \columnwidth]{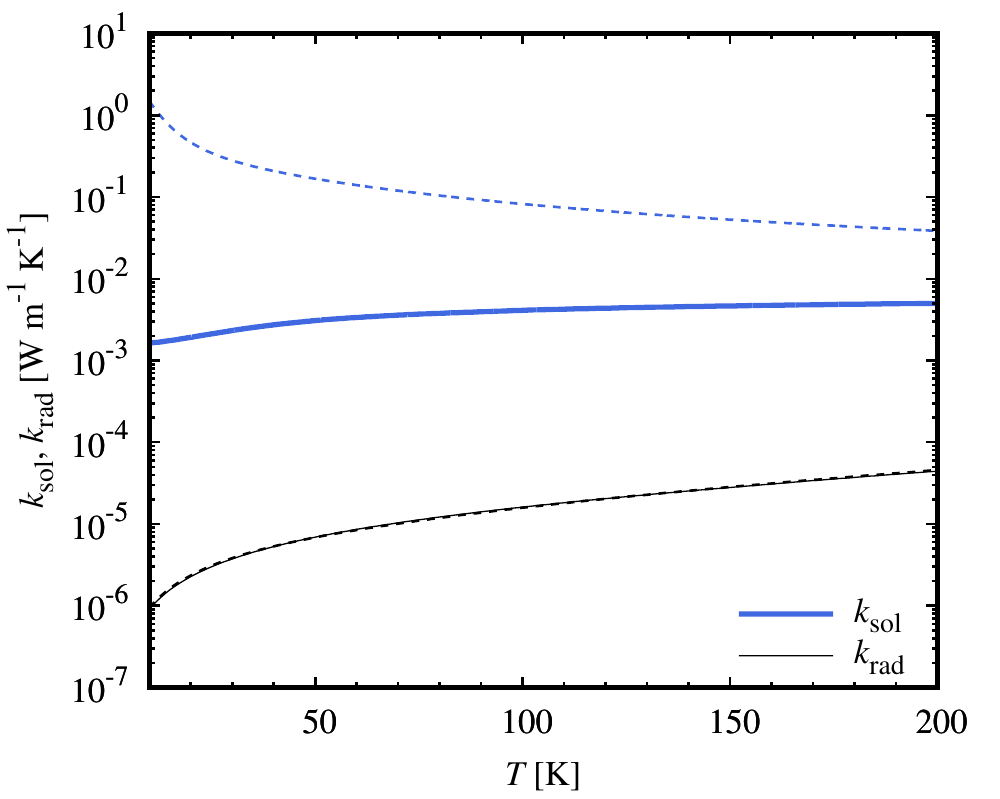}
\caption{
Thermal conductivity through the the solid network, $k_{\rm sol}$ (blue lines), and the thermal conductivity due to radiative transfer within a pebble, $k_{\rm rad}$ (black lines).
The solid lines represent the case of organic--silicate grains, whereas the dashed lines represent the case of ice--organic--silicate grains.
}
\label{fig.app.krad}
\end{figure}

Figure \ref{fig.app.krad} shows the two terms of the thermal conductivity of the constituent aggregates, i.e., $k_{\rm sol}$ and $k_{\rm rad}$.
We confirmed that $k_{\rm rad}$ is several orders of magnitude lower than $k_{\rm sol}$, and the thermal conductivity of the pebbles is $k_{\rm agg} \simeq k_{\rm sol}$, as was assumed in Section \ref{sec.k.agg}.

\section{Effective absorption cross section of pebbles}
\label{app.sigma}

In Appendix \ref{app.sigma}, we show that the absorption cross section can be approximated by the geometric cross-section for the pebble sizes examined in this study.
The effective absorption cross sections of the pebbles were calculated as
\begin{equation}
\sigma_{\rm eff, agg} = M_{\rm agg} \kappa_{\rm eff, agg},
\end{equation}
where $M_{\rm agg} = 4 \pi \rho_{\rm m} \phi_{\rm agg} {R_{\rm agg}}^{3} / 3$ is the mass of the pebbles and $\kappa_{\rm eff, agg}$ is the effective absorption coefficient of the pebbles.
We adopt the method used in Appendix \ref{app.rad}, but in addition, we consider the presence of voids ($\epsilon_{\rm void} = 1$) in Equation (\ref{eq:bruggeman}) to calculate the effective dielectric function.

\begin{figure}
\centering
\includegraphics[width = \columnwidth]{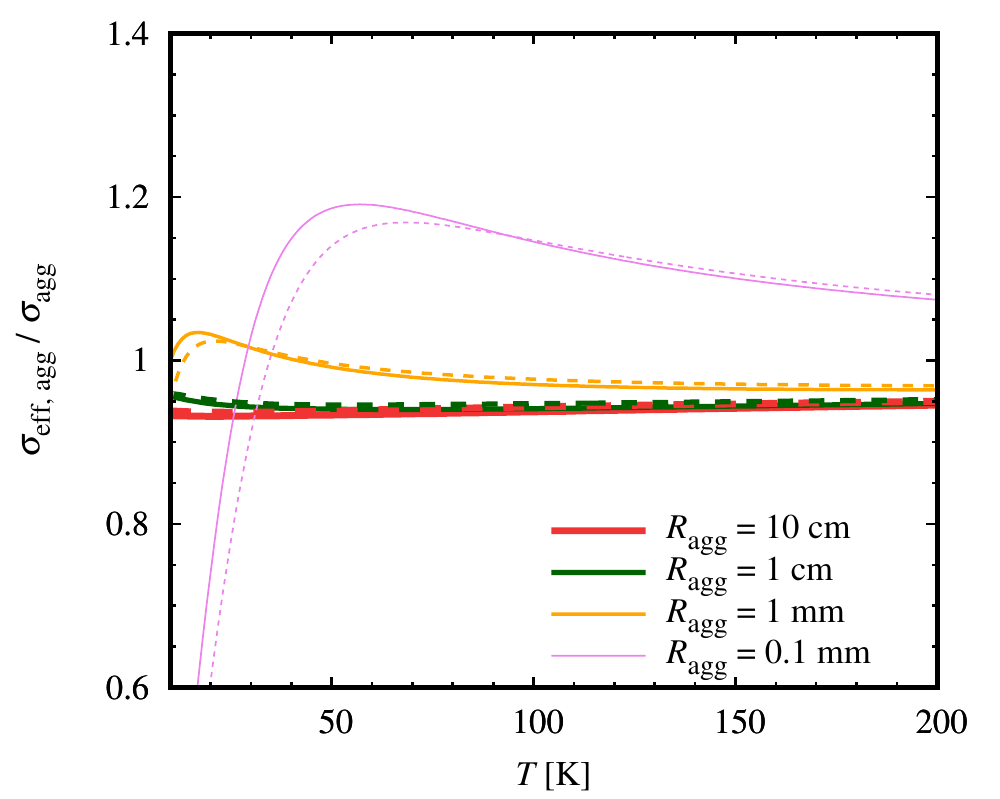}
\caption{
Effective absorption cross sections of the pebbles ($R_{\rm agg} = 0.1\ {\rm mm}$--$10\ {\rm cm}$).
The solid lines represent the case of organic--silicate grains, and the dashed lines represent the case of ice--organic--silicate grains.
}
\label{fig.app.cs}
\end{figure}

Figure \ref{fig.app.cs} shows the effective absorption cross sections of pebbles that are normalized by the geometric cross section, $\sigma_{\rm eff, agg} / \sigma_{\rm agg}$.
We found that, in the temperature range of $T \gtrsim 30\ {\rm K}$, the normalized cross section is in the range of $0.9 \lesssim \sigma_{\rm eff, agg} / \sigma_{\rm agg} \lesssim 1.2$ for the aggregate radius of $R_{\rm agg} \ge 0.1\ {\rm mm}$.
Then we can approximate $\sigma_{\rm eff, agg}$ as $\sigma_{\rm agg}$ as shown in Section \ref{sec.k.hie}.

\section{Diurnal and orbital skin depths}
\label{app.ds}

\begin{figure}
\centering
\includegraphics[width = \columnwidth]{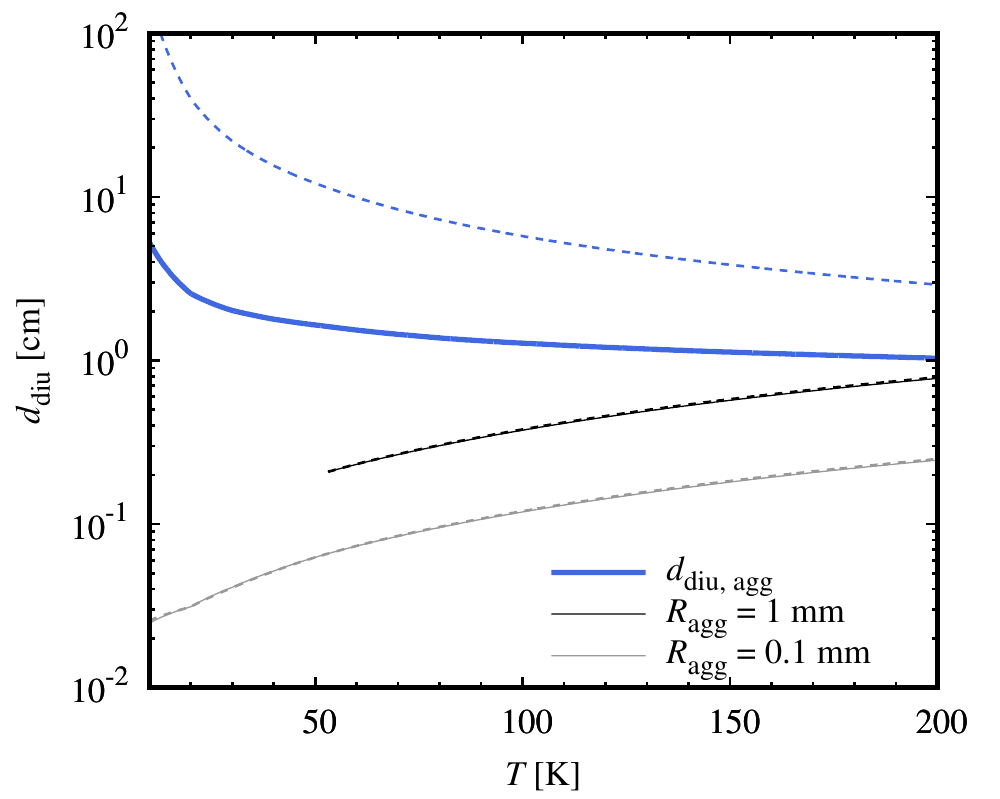}
\caption{
Diurnal thermal skin depth as a function of temperature.
The blue lines represent the diurnal thermal skin depth of the constituent aggregates, $d_{\rm diu, agg}$.
The black lines represent the diurnal thermal skin depth of the hierarchical aggregates, $d_{\rm diu, hie}$, for the case of $R_{\rm agg} = 1\ {\rm mm}$, and the grey lines represent $I_{\rm hie}$ for the case of $R_{\rm agg} = 0.1\ {\rm mm}$.
The solid lines represent the case of organic--silicate grains, whereas the dashed lines indicate the case of ice--organic--silicate grains.
}
\label{fig.app.diu}
\end{figure}

\begin{figure}
\centering
\includegraphics[width = \columnwidth]{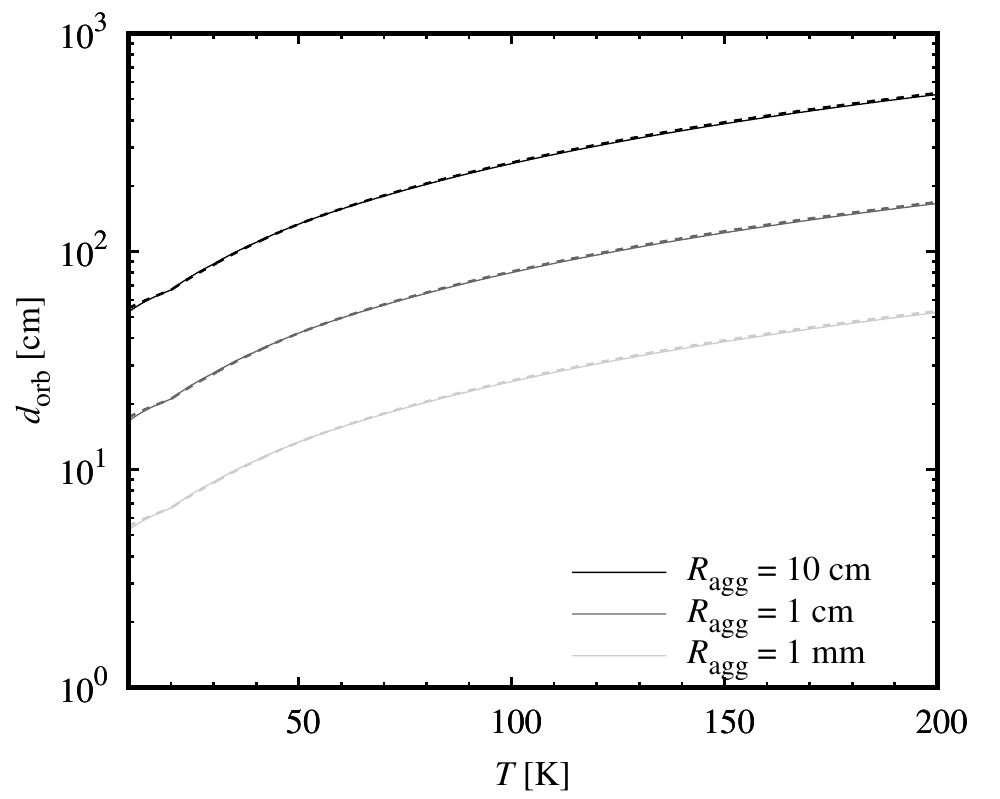}
\caption{
Orbital thermal skin depth as a function of temperature.
The color of the lines indicate the aggregate radius ($R_{\rm agg} = 10\ {\rm cm}$, $1\ {\rm cm}$, and $1\ {\rm mm}$).
In these cases, the orbital thermal skin depth is given by $d_{\rm orb} = d_{\rm orb, hie}$.
The solid lines are for the case of organic--silicate grains, and the dashed lines are for the case of ice--organic--silicate grains.
}
\label{fig.app.orb}
\end{figure}

In Appendix \ref{app.ds}, we show the diurnal and orbital thermal skin depths.
Figures \ref{fig.app.diu} and \ref{fig.app.orb} are the diurnal and orbital thermal skin depths, $d_{\rm diu}$ and $d_{\rm orb}$, as functions of temperature.
We note that the orbital thermal skin depth is larger than the aggregate radius for the case of $R_{\rm agg} \lesssim 1\ {\rm m}$.

\bsp	
\label{lastpage}
\end{document}